\documentclass[%
reprint,
superscriptaddress,
amsmath,amssymb,
aps,
prb,
]{revtex4-2}

\usepackage{graphicx}
\usepackage{dcolumn}
\usepackage{bm}

\usepackage[utf8]{inputenc}
\usepackage[T1]{fontenc}
\usepackage{booktabs, array, mathptmx, float, tabularx, booktabs, lipsum, amsmath,multirow}
\usepackage{siunitx, xcolor}
\usepackage[version=4]{mhchem}

\usepackage{times}
\usepackage{amsmath}
\usepackage{mathptmx}
\DeclareMathAlphabet{\mathcal}{OMS}{cmsy}{m}{n}
\DeclareSymbolFont{largesymbols}{OMX}{cmex}{m}{n}
\usepackage{mathrsfs}
\usepackage{epstopdf}
\usepackage[colorlinks=true, letterpaper=ture, pdfstartview=FitV, linkcolor=blue, citecolor=blue, urlcolor=blue]{hyperref}
\usepackage{dcolumn}

\begin{document}
	
	\title{Controllable Creation of Skyrmion Bags in a Ferromagnetic Nanodisk}%
	
	\author{Lan Bo}%
	\affiliation{Key Laboratory for Anisotropy and Texture of Materials (MOE), School of Materials Science and Engineering, Northeastern University, Shenyang 110819, China}
	\affiliation{Institute of Advanced Magnetic Materials, College of Materials and Environmental Engineering, Hangzhou Dianzi University, Hangzhou 310012, China}
	\affiliation{Department of Applied Physics, Waseda University, Okubo, Shinjuku-ku, Tokyo 169-8555, Japan}
	\author{Rongzhi Zhao}%
	\affiliation{Key Laboratory for Anisotropy and Texture of Materials (MOE), School of Materials Science and Engineering, Northeastern University, Shenyang 110819, China}
	\affiliation{Institute of Advanced Magnetic Materials, College of Materials and Environmental Engineering, Hangzhou Dianzi University, Hangzhou 310012, China}
	\author{Chenglong Hu}%
	\affiliation{Key Laboratory for Anisotropy and Texture of Materials (MOE), School of Materials Science and Engineering, Northeastern University, Shenyang 110819, China}
	\affiliation{Institute of Advanced Magnetic Materials, College of Materials and Environmental Engineering, Hangzhou Dianzi University, Hangzhou 310012, China}
	\author{Xichao Zhang}%
	\affiliation{Department of Applied Physics, Waseda University, Okubo, Shinjuku-ku, Tokyo 169-8555, Japan}
	\author{Xuefeng Zhang}%
	\email[Corresponding E-mail: ]{zhang@hdu.edu.cn}
	\affiliation{Key Laboratory for Anisotropy and Texture of Materials (MOE), School of Materials Science and Engineering, Northeastern University, Shenyang 110819, China}
	\affiliation{Institute of Advanced Magnetic Materials, College of Materials and Environmental Engineering, Hangzhou Dianzi University, Hangzhou 310012, China}
	\author{Masahito Mochizuki}%
	\affiliation{Department of Applied Physics, Waseda University, Okubo, Shinjuku-ku, Tokyo 169-8555, Japan}
	
	\date{\today}%
	
	\begin{abstract}
		Skyrmion bags are composed of an outer skyrmion and arbitrary inner skyrmions, which have recently been observed in bulk chiral magnets, but still remain elusive in magnetic films. Here, we propose a method of creating skyrmion bags in a thin-film nanodisk, which includes three steps. Firstly, the size of outer skyrmion is enlarged by a vertical magnetic field, then inner skyrmions are nucleated at an off-center area by local current injection, and the system is finally reconstructed due to multiple inter-skyrmion potentials. Thus, skyrmion bags with topological charge up to forty can be created. Simulated Lorentz transmission electron microscopy images are given to facilitate the experimental demonstration. Our proposal is expected to inspire relevant experiments in magnetic films, and pave the way for potential spintronic applications based on skyrmion bags.
	\end{abstract}
	
	\maketitle
	
	\section{Introduction}
	
	The inter-discipline of topological textures and spintronics has attracted considerable interest and attention \cite{mochizuki2015dynamical,zhang2020skyrmion,bo2022micromagnetic} since the first experimental observation of magnetic skyrmions \cite{yu2010real}. Many non-collinear spin textures as variations and extensions of skyrmions have also been predicted and observed, some of which promise even greater advantages compared to conventional skyrmions \cite{gobel2021beyond}. In 2019, skyrmion bags were discovered experimentally in liquid crystals \cite{foster2019two}, and then predicted to exist in magnetic systems \cite{foster2019two,rybakov2019chiral}. The structure of such topological textures seems like an outer skyrmion (oSk) bagging multiple small inner skyrmions (iSks) with opposite polarity. Attributed to the arbitrary topological charge, they are expected to extend additional degree of freedom for data encoding in skyrmion-based racetrack memory. Therefore, intensive studies were reported successively in terms of their existence and stability \cite{kind2020existence}, spin excitation modes \cite{zeng2022spin}, and dynamical behaviors driven by spin-orbit torque \cite{zeng2020dynamics}, spin-transfer torque \cite{kind2021magnetic}, and anisotropy gradient \cite{zeng2022skyrmion}.
	
	Similar to conventional magnetic skyrmions, skyrmion bags are believed to exist in both chiral magnets with bulk Dzyaloshinskii-Moriya interaction (DMI) \cite{dzyaloshinsky1958thermodynamic,moriya1960anisotropic}, and magnetic films with interfacial DMI. On the one hand, chiral skyrmions are prone to form chain states \cite{du2015edge,leonov2022skyrmion} or cluster states \cite{zhang2017skyrmion,jiang2018dynamics,leonov2019skyrmion,zheng2021magnetic}, so skyrmion bags in B20-type magnets have been observed very recently \cite{tang2021magnetic}, although they were termed as skyrmion bundles due to morphological distortion in three-dimensional (3D) space. On the other hand, skyrmion bags in magnetic films have yet been confirmed in experiments. However, remarkedly, theoretical studies \cite{kind2020existence,zeng2022spin,zeng2020dynamics,kind2021magnetic,zeng2022skyrmion} were all carried out in two-dimensional (2D) systems, and their initial configurations of skyrmion bags were all set artificially. Therefore, it is in an urgent need to explore an effective method for the creation of skyrmion bags in magnetic films to bridge the gap between experimental and theoretical results. 
	
	Pervious works have shown that skyrmions could be generated by local current injection \cite{tchoe2012skyrmion,iwasaki2013current,jiang2015blowing,lin2016edge,heinonen2016generation,wang2022electrical}. But even with similar methods, more elaborate design is needed to create skyrmion bags owing to their complex topology. Herein, we propose an available method to achieve highly controllable creation of skyrmion bags in a 2D magnetic nanodisk. The process can be vividly summarized as three steps: first loosen the oSk, then put in an iSk, and finally tighten the oSk. The first and second steps are controlled by a vertical magnetic field and local spin currents, respectively, and the third step is realized with the assistance of the multiple skyrmion-skyrmion interactions. Thus, skyrmion bags with arbitrary topological charge could be created. 
	
	\section{Model and methodology}
	
	In the 2D model, the topological charge number $Q$ is a classical parameter to describe a spin texture, defined as \cite{nagaosa2013topological}
	\begin{align}
		{Q}=\frac{1}{4\pi}\int{\bf m}\cdot \left(\frac{\partial {\bf m}}{\partial x} \times \frac{\partial{\bf m}}{\partial y}\right) \,{\rm d}x\,{\rm d}y
		,\label{1}
	\end{align}
	where ${\bf m}$ is the unit vector of magnetization. Although skyrmion bags can also be described by $Q$, here we use a simpler notation $S(Q+1)$ \cite{foster2019two,kind2020existence,kind2021magnetic,zeng2022skyrmion} because there are overall $Q+1$ nested iSks. For example, as shown in Fig.~\ref{1} (a), skyrmion bags with $Q=2$ have three iSks, thus labeled as $S(3)$. A single skyrmion with $Q=-1$ \cite{bogdanov1999stability} and a skyrmionium with $Q=0$ \cite{zhang2016control} are marked by $S(0)$ and $S(1)$, respectively.
	
	As shown in Fig.~\ref{1} (b), the model used for micromagnetic simulations is a bilayer disk of Co (1 nm)/Pt (3 nm) with 1024 nm in diameter. An off-center nanocontact with diameter $\Phi$ is placed along the positive direction of the $x$-axis, 128 nm away from the edge. A similar geometry has been reported effective to create a single skyrmion \cite{durrenfeld2017controlled}. When vertical currents inject into the nanocontact, it can act as a skyrmion generator induced by spin Hall effect (SHE) \cite{durrenfeld2017controlled} or spin transfer torque (STT)  \cite{liu2015switching}. Additionally, an external field ${\bf H}_{\rm ext}=(0,0,H_z)$ is applied in the whole space.
	
	\begin{figure}[t]
		\includegraphics[width=1\linewidth]{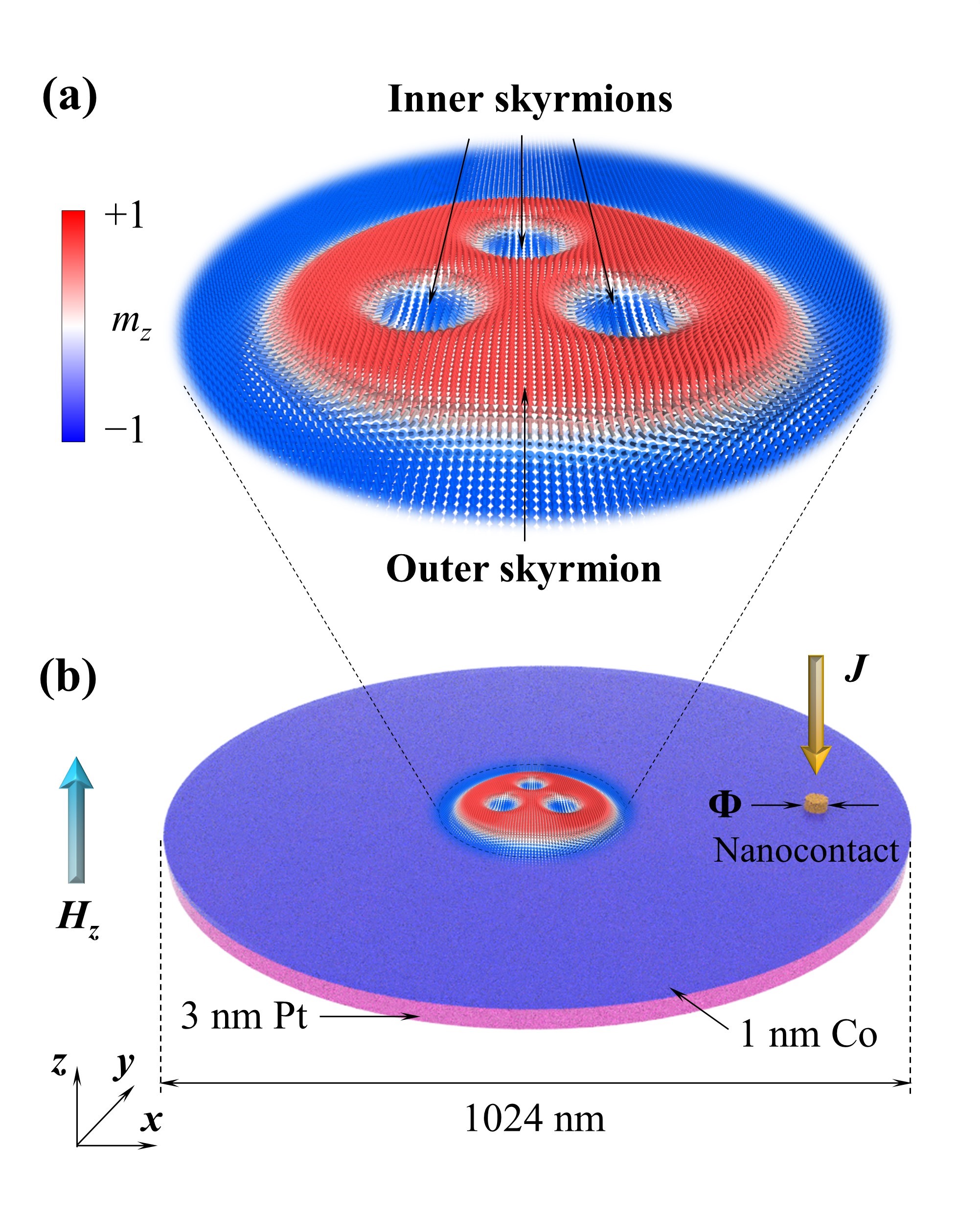}
		\caption{(a) Spin configuration of stable skyrmion bag $S(3)$ with topological charge number $Q=2$. It is composed of an outer skyrmion (oSk) and three inner skyrmions (iSks) with opposite polarity. The color bar of $m_z$ is also applicable to all the spin configurations presented below. (b) Schematic of the micromagnetic model. The diameter of the nanodisk is 1024 nm, and the thickness is 4 nm, including 3-nm-Pt and 1-nm-Co. A nanocontact with diameter $\Phi$ is positioned in the positive direction of the $x$-axis, 128 nm from the edge. The vertical current $J$ is applied locally to the nucleation area, and the vertical magnetic field $H_z$ is applied to the entire nanodisk.}\label{1} 
	\end{figure}
	
	The total energy of this system is given by 
	\begin{align}
		E&=\int{\rm d} {\bf r} \bigg\{A(\nabla {\bf m})^2+D\left[m_{z}(\nabla \cdot {\bf m})-({\bf m} \cdot \nabla) m_z\right]
		\nonumber\\
		&\quad +K m_z^2+\mu_0 M_{\rm s} m_z H_z-\frac{1}{2} \mu_0 M_{\rm s} {\bf m} \cdot {\bf H}_{\rm dm}\bigg\}
		,\label{2}		
	\end{align}
	where $A$ is the Heisenberg exchange constant, $D$ is the antisymmetric interfacial DMI constant, $K$ is the perpendicular anisotropy constant, $\mu_0$ is the vacuum permeability, $M_{\rm s}$ is the saturation magnetization, $H_z$ is the static Zeeman ﬁeld applied along the $z$-axis, and ${\bf H}_{\rm{dm}}$ is the demagnetizing field. To describe the static states and dynamical behaviors of skyrmion bags, we exploit the Landau--Lifshitz--Gilbert (LLG) equation with Slonczewski STT term \cite{slonczewski1996current}:
	\begin{align}
		\frac{\partial {\bf m}}{\partial t}&=-\gamma {\bf m} \times {\bf H}_{\rm eff}+\alpha\left({\bf m} \times \frac{\partial {\bf m}}{\partial t}\right)+\tau_{\rm STT}
		,\label{3} \\
		\tau_{\rm STT}&=\frac{\hbar \gamma P J}{M_{\rm s} e d}[\,{\bf m} \times({\bf m}_p \times {\bf m})]
		,\label{4}
	\end{align}
	where $ {\bf H}_{\rm eff}=-(\delta \epsilon / \delta {\bf m}) /(\mu_0 M_{\rm s})$ is the effective field, with $\epsilon$ denoting the average energy density determined from Eq. (\ref{2}). $J$ is the current density along the $z$-axis, $P$ is the degree of spin polarization, ${\bf m}_{p}$ is the unit vector of the polarization direction. $\hbar$, $e$, and $d$, are the reduced Planck’s constant, the elementary electron charge, and the thickness of ferromagnetic layer, respectively.
	
	To solve Eq. (\ref{3}) \& (\ref{4}), we perform micromagnetic simulations using the MuMax3 finite-difference GPU accelerated code \cite{vansteenkiste2014design}. The ferromagnetic layer is discretized into $1024\times1024\times1$ cubes with a side length of 1 nm, less than the magnetocrystalline exchange length $\sqrt{ A / K }$ and the magnetostatic exchange length $\sqrt{2 A / (\mu_0 M_{\mathrm{s}}^2)}$ \cite{abo2013definition}. Material parameters are derived from Co/Pt films in real experiments \cite{sampaio2013nucleation,metaxas2007creep}, which has also been proved to be capable of hosting skyrmion bags \cite{kind2020existence}: $M_{\rm s}=5.8\times10^{5}\, \rm{A}/\rm{m}$,
	$A=1.5\times10^{-11}\ \rm{J}/\rm{m}$,
	$D=3.5\times10^{-3}\ \rm{J}/\rm{m}^2$,
	$K=8.0\times10^5\ \rm{J}/\rm{m}^3$, and $\alpha=0.3$. Key inputs of Slonczewski STT are $P=0.4$ and ${\bf m}_{p}=(0,0,1)$ \cite{liu2015switching}. External excitations of $H_z$ and $J$ are varied to control the skyrmion bags. All simulations are performed at temperature $T=0\ \rm{K}$ in the main text.

	\section{Results and discussion}
	We start by determining the equilibrium states and external field dependence of the system, which is preparation for regulating the size of the oSk. The radius of an equilibrium skyrmion under zero external field can be estimated by \cite{rohart2013skyrmion}
	\begin{align}
		r_{\rm sk} \approx \frac{\sqrt{A / K_{\rm eff}}}{\sqrt{2\left(1-D / D_{\rm c}\right)}}
		,\label{5}
	\end{align}
	where $K_{\rm eff}=K-\mu_0 M_{\mathrm{s}}^2 / 2$ is the effective anisotropy constant, and $ D_{\rm c}=4 \sqrt{A \cdot K_{\mathrm{eff}}} / \pi$ is the critical DMI value. For the present system,  $r_{\rm sk}$ is calculated to be 13.0 nm, so a single skyrmion with positive polarity and $r_{\rm sk}=13.0\ \rm{nm}$ is chosen as the initial state. The conjugate gradient method is used to find the energy minimum within the range of $H_z=\pm30\ \rm{mT}$. As shown in Fig.~\ref{2} (a), the simulation result of $r_{\rm sk}$ under zero field is 16.5 nm, which is very close to the theoretical value. For $H_z<13\ \rm{mT}$, the skyrmion expands slightly with the increase of $H_z$ and almost coincides with the infinite film solution \cite{rohart2013skyrmion}. When $H_z\approx13 \ \rm{mT}$, the skyrmion enlarges dramatically to become a circular domain with the size comparable with the nanodisk. As $H_z$ increases further, $r_{\rm sk}$ no longer grows up sharply owing to the edge confinement. Similar results have also been discussed in Refs. \cite{tejo2018distinct,tomasello2018origin}. To find the threshold $H_z$, we also calculate the total energy $E$ of skyrmion states, and two quasi-uniform states with $m_z=\pm1$ as references, as shown in Fig.~\ref{2} (b). It can be seen that the energy of skyrmion is only slightly higher than that of quasi-uniform state under negative $H_z$; while for positive $H_z$, skyrmion becomes metastable, and it is relatively more stable when $H_z>13\ \rm{mT}$. The above results provide the threshold, which means that $H_z$ that is greater than or less than 13 mT can control the skyrmion bags loosening or tightening. In the following, $H_z=20\ \rm{mT}$ and $H_z=-10\ \rm{mT}$ are chosen as typical cases to study.
	
	\begin{figure}[t]
		\includegraphics[width=1\linewidth]{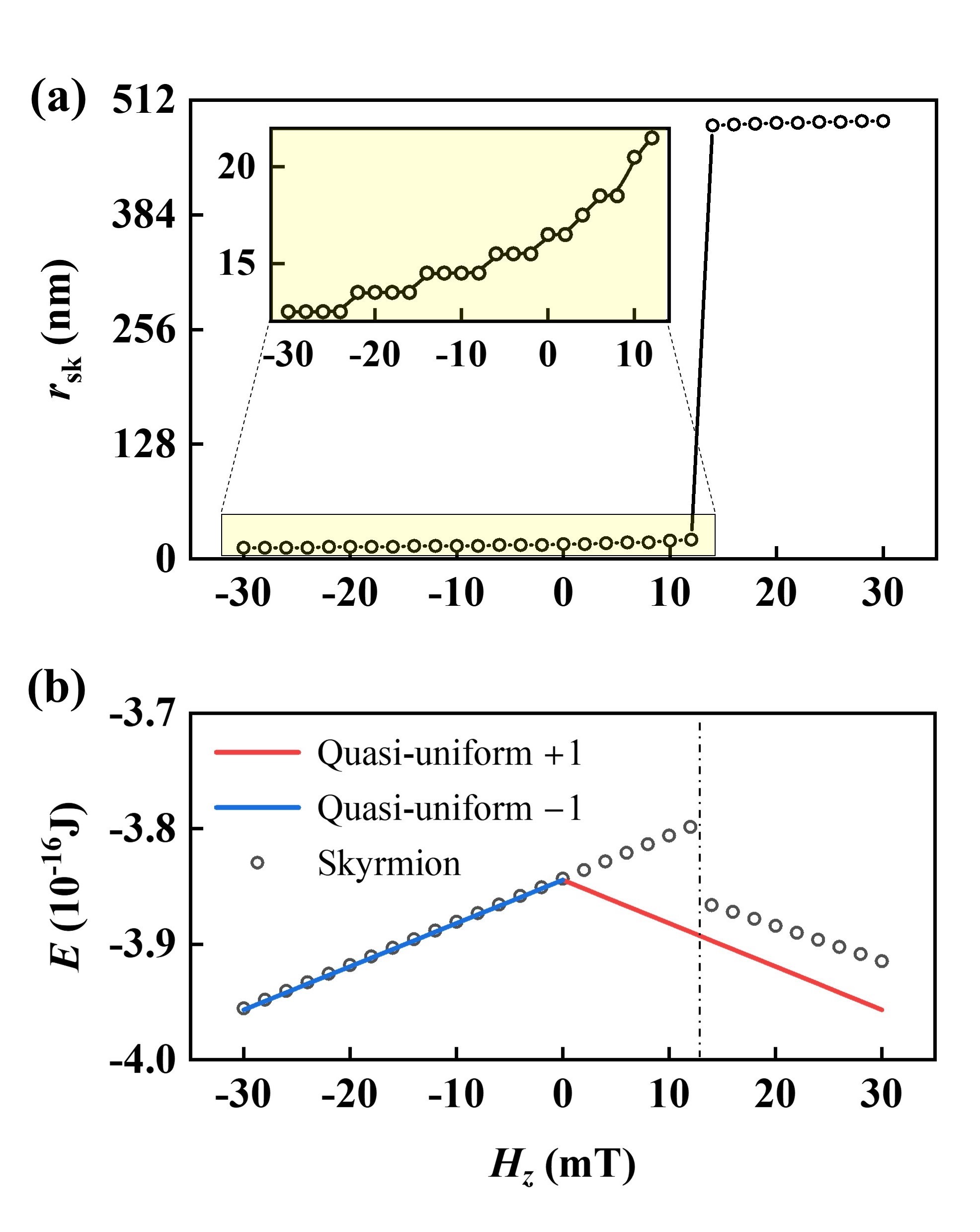}
		\caption{External field dependence of a single skyrmion with positive polarity on the nanodisk. (a) Variation of the skyrmion radius $r_{\rm sk}$ versus the vertical magnetic field $H_z$. The inset shows zoom-in plot for $H_z<13\  \rm{mT}$. (b) Variation of the system energy $E$ versus $H_z$, for different equilibrium states (quasi-uniform magnetizations and a single skyrmion). Vertical chain line marks the sudden change of $E$.}\label{2}
	\end{figure}
	
	Next, let us focus on the nucleation of an iSk. During this process, $H_z=20\ {\rm mT}$ is applied to keep the skyrmion bags loosening, so that the iSk can be nucleated inside the oSk. Herein, a vertical current is applied locally to the nanocontact, which acts as a spin polarizer to provide STT effect \cite{liu2015switching}. Note that a similar method induced by SHE is also an equivalent candidate \cite{durrenfeld2017controlled}. We first inject a constant current with the intensity $I$ into the nanocontact with the diameter $\Phi$. A phase diagram with respect to $I$ and $\Phi$ is mapped to describe whether a stable iSk can exist or not. As shown in Fig.~\ref{3} (a), the stable (green) region and unstable (red) region have obvious boundary. With the increase of $\Phi$, the $I$ value required to generate stable iSk decreases first and then increases. Considering that $r_{\rm sk}$ under zero external field is about 13.0 nm to 16.5 nm [see Eq. (\ref{5}) and Fig.~\ref{2} (a)], we hereby fix $\Phi$ at an equivalent value of 30 nm for following studies. Next, in order to give insight into the stability mechanism of the nucleated iSk, we introduce the system maximum torque $Tor_{\rm max}={\rm max}(d{\bf m}/dt)$ over all cells, and plot its variation versus time in Fig.~\ref{3} (b). Obviously, $Tor_{\rm max}$ has two peaks for stable situation ($I=20,25\ {\rm mA}$), but has only one peak for unstable one ($I=10,15\ {\rm mA}$) even if the constant current continues for 100 ps. The first peak represents the appearance of reverse spins, while the second peak represents that those spins overcome a local twist and arrange into a Néel-type skyrmion. If $I$ is too small to rearrange the twist, the unstable reverse spins will annihilate at last. Snapshots of local spin configurations at 50 ps for the two situations are shown in the insets of Fig.~\ref{3} (b), where the abovementioned twist is marked by a yellow circle. In the following, $I=25\ {\rm mA}$ is chosen for creating iSks, which is close to the value of 22 mA reported in Ref. \cite{durrenfeld2017controlled}. To resemble an experimentally achievable generator output, the current is applied in the form of a picosecond pulse, as shown in the inset of Fig.~\ref{3} (b). The pulse is consisted of 10 ps each rising\,/\,falling time and 50 ps constant current, which is enough for $I=25\ {\rm mA}$ to overcome the local twist.  Additionally, the nucleation is also infulenced by other factors, such as discretization of the simulations, damping, finite temperature, and the shape of the current pulse. See results and discussion about those factors in the Supplemental Material [URL will be inserted by the publisher].
	
	Although an iSk is created now, it is located underneath the nanocontact, which will cause trouble for the nucleation of subsequent iSks. So, in this step, we aim at moving the iSk to the central area of the nanodisk. As has been demonstrated, the skyrmion bags will tighten when $H_z<13\ \rm{mT}$, so it is expected that the iSk will be pushed toward the center by the interaction potential from the oSk. To accelerate this process, $H_z=-10\ {\rm mT}$ is applied, and the snapshots of dynamical spin configurations at selected times are shown in Fig.~\ref{3} (d). When $t=2\ {\rm ns}$, the outer circular chiral domain wall of the oSk contacts the iSk and produces slight deformation; the iSk hereby starts moving and then almost arrives at the central area at 12 ns. After the system is fully stabilized at 14 ns, $H_z=20\ {\rm mT}$ is applied to loosen the skyrmion bags again with the iSk being left at the center, so that the nucleation of next iSk will not be influenced. It is worth noting that the motion of iSk toward the center does not follow a straight line, but a curve shown in Fig.~\ref{3} (e). Assuming that the iSk has a rigid structure, the motion trajectory can be understood by Thiele framework \cite{thiele1973steady}:
	\begin{align}
		{\bf G} \times {\dot{\bf R}} -\alpha \mathcal{D} \cdot {\dot{\bf R}} + {\bf F}_{\rm p}({\bf R})= {\bf 0}
		,\label{6}
	\end{align}
	where $ {\bf G}=G\hat{z} $ is the gyromagnetic coupling vector that relates to $Q$ defined in Eq. (\ref{1}), $\mathcal{D}$ is the dissipative tensor, ${\dot{\bf R}}$ is the motion velocity, and ${\bf R}$ is the position $(x, y)$ of the nucleated iSk, defined by the cells with magnetization component $m_{z}=-1$. The first term is the Magnus force ${\bf F}_{\rm g}$, the second term is the dissipative force ${\bf F}_{\rm d}$, and the third term ${\bf F}_{\rm p}$ is the force due to the confining potential. Similar confining potential has been reported effective in regulating skyrmion motion \cite{xing2020enhanced,yang2022inhibition}, which is induced by a DMI-determined twist of the magnetization \cite{zhang2015skyrmion,zhang2017motion,bo2022velocity}. A zoom-in picture of the spin configuration selected at 7 ns is shown in Fig.~\ref{3} (c), with a schematic of instantaneous force analysis. At this moment, ${\dot{\bf R}}$ is nearly along the $-y$ direction. Because ${\bf F}_{\rm g}$ and ${\bf F}_{\rm d}$ are always perpendicular or antiparallel to ${\dot{\bf R}}$, they are currently along the $-x$ and $+y$ direction, respectively. ${\bf F}_{\rm p}$ is always perpendicular to the tangent of oSk and iSk, which play the role of balancing ${\bf F}_{\rm g}$ and ${\bf F}_{\rm d}$, and also providing the driving force for iSk motion. The competition among the above forces jointly leads to the curve trajectory shown in Fig.~\ref{3} (e). 
	
	\begin{figure*}[t]
		\includegraphics[width=1\linewidth]{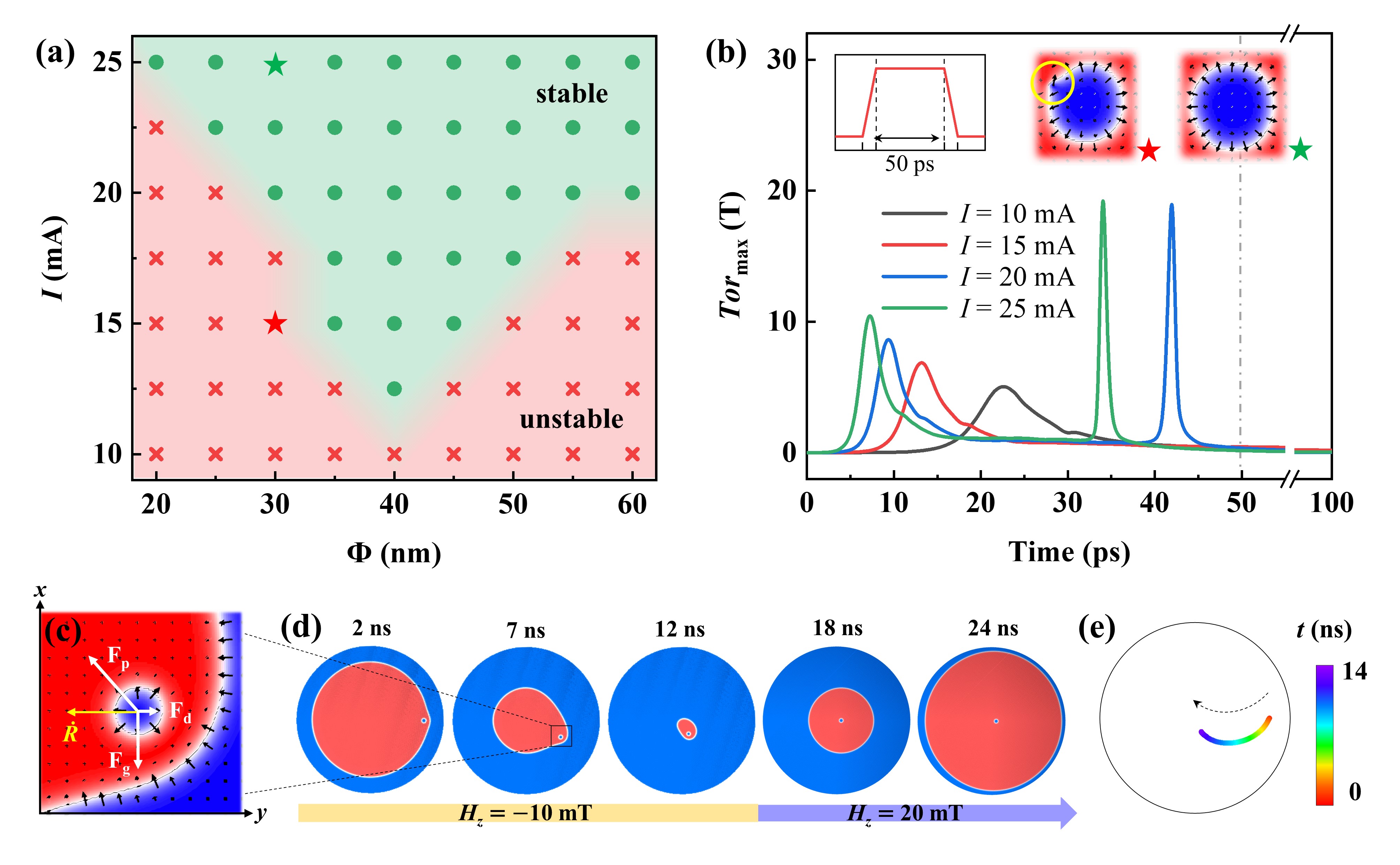}
		\caption{(a) Phase diagram of the nucleated iSk stability with respect to various nanocontact diameter $\Phi$ and current intensity $I$. Green represents stable situation and red represents unstable one. The star symbols correspond to the local spin configurations shown in the insets of Fig.~\ref{3} (b). (b) The system maximum torque $Tor_{\rm max}$ plotted as a function of time for various $I$. Left inset: temporal shape of the imposed pulse with 50 ps constant current. Right insets: snapshots of local spin configurations at 50 ps for unstable\,/\,stable iSk, with a yellow circle marking the local twist. (c) Zoom-in picture of local spin configuration selected at 7 ps, and the schematic of force analysis. The white arrows only indicate the direction of the forces, not the realistic ratio of amplitudes. (d) Snapshots of dynamical spin configurations at selected times during the period of skyrmion bags tightening ($H_z=-10\ {\rm mT}$) and loosening again ($H_z=20\ {\rm mT}$). (e) Time-dependent motion trajectory of iSk during the skyrmion bags tightening. The arrow marks the direction of the motion.}\label{3} 
	\end{figure*}
	
	So far, we have presented the whole cycle of the creation of $S(1)$. By repeating this cycle, skyrmion bags with higher topological charge can be obtained. In Figs.~\ref{4} (a)--(c), the three steps of the creation are shown, from $S(1)$--$S(2)$ to $S(15)$--$S(16)$. Fig.~\ref{4} (a) shows the loosened states of skyrmion bags under $H_z=20\ {\rm mT}$, Fig.~\ref{4} (b) shows the spin configurations when an iSk is just nucleated, selected after the current pulse being applied, and Fig.~\ref{4} (c) shows the stable states of skyrmion bags. To further bring the theoretical results closer to real experiments, we give the simulated Lorentz transmission electron microscopy (L-TEM) images \cite{walton2013malts,mccray2021understanding}. This approach is usually adopted for complex topological configurations that have not been observed experimentally, such as magnetic hopfions \cite{tai2018static,voinescu2020hopf,bo2021spin}. When considering the real application of the proposed system, some amount of capping layers is necessary to provide electric connection and insulate surrounding to the nanocontact, which may influence the direct imaging underneath the nanocontact. So, we only show the partly enlarged images of the stable skyrmion bags in Fig.~\ref{4} (d). Assuming that full electron-wave processing of the electron beam is used within the small defocus limit, the L-TEM contrast generated by the underlying skyrmion bags can be described by the curl of the magnetization along the beam propagation axis ${\bf \hat{z}}$, given by \cite{pollard2017observation}
	\begin{align}
		I({\bf R},\Delta)=1-(\Delta e \mu_0 \lambda d / \hbar)(\nabla\times{\bf R})\cdot{\bf \hat{z}}
		,\label{7}
	\end{align}
	where $I$ is the normalized intensity, $\Delta$ is the degree of defocus, and $\lambda$ is the electron wavelength. Here, ${\bf \hat{z}}$ is tilted to $20^\circ$ to simulate the tilt of the sample in real experiments. These zoom-in images clearly depict the morphology of skyrmion bags, which can be compared with skyrmions that already observed in Co-based thin films \cite{pollard2017observation,he2017realization,lin2018observation} and facilitate the experimental demonstration of our simulation results.
	
	\begin{figure*}[t]
		\includegraphics[width=1\linewidth]{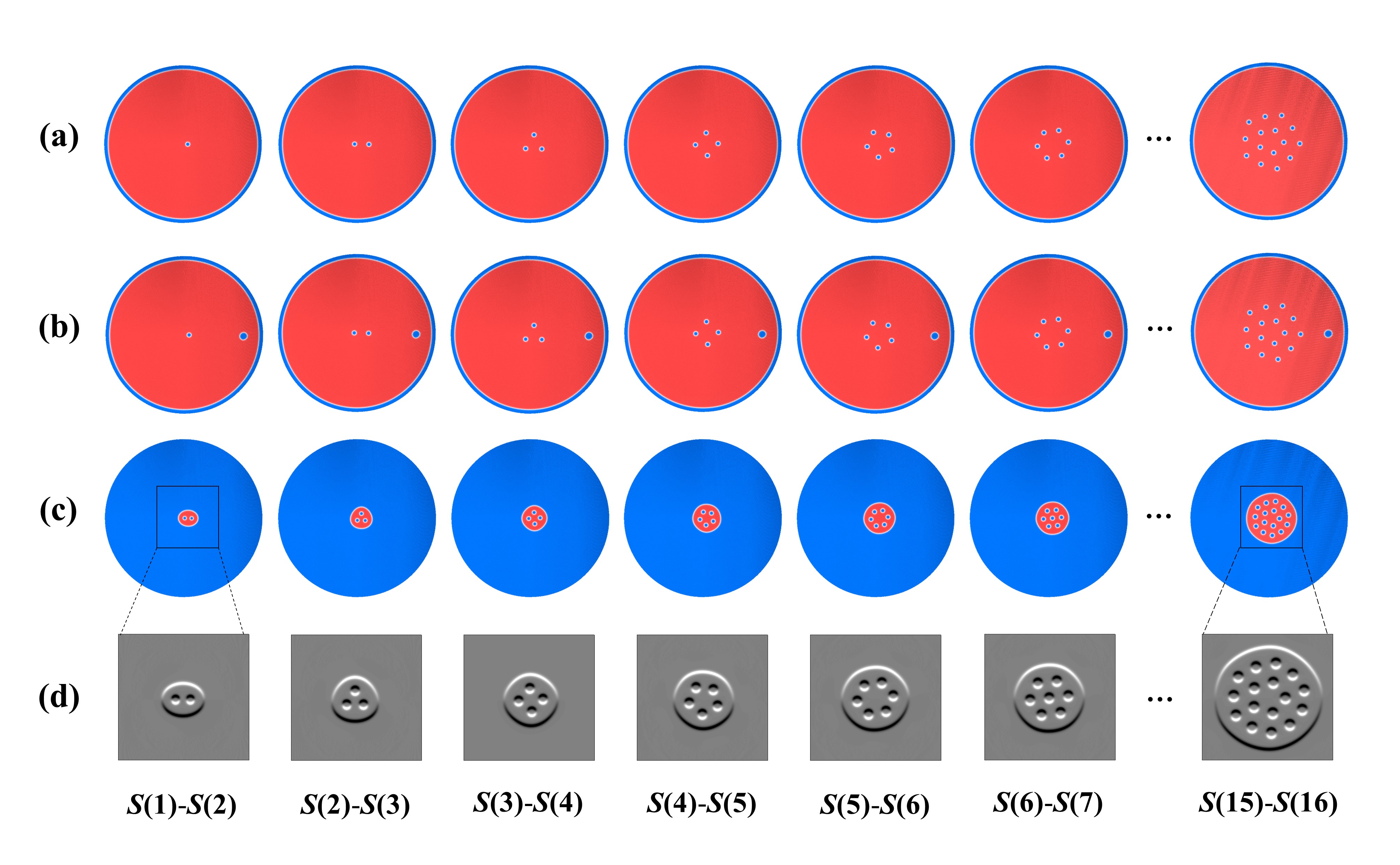}
		\caption{Spin configurations of $S(1)$--$S(2)$ to $S(15)$--$S(16)$ for the three steps of creating skyrmion bags: (a) the loosened states of skyrmion bags under $H_z=20\ {\rm mT}$; (b) the states that an iSk is just nucleated, selected after the  current pulse being applied; (c) the stable tightened states of skyrmion bags under $H_z=0\ {\rm mT}$. (d) Zoom-in images of the stable skyrmion bags processed by computer-simulated Lorentz transmission electron microscopy. The contrast is derived from the curl of the magnetization along the beam propagation axis $(0, \arcsin20^\circ, \arccos20^\circ)$.}\label{4}     
	\end{figure*}
	
	Finally, we investegate the differences in skyrmion bags with various topological charge, in terms of their dynamics and static properties. In Figs.~\ref{5} (a) (b), the chosen external field $H_z=20\ {\rm mT}$ and $H_z=-10\ {\rm mT}$ are applied for $S(1)$ to $S(16)$, to study the time-dependent radius variation of oSks. Here, third-order Runge-Kutta method is used to run the calculation, and the results are also verified by fifth-order solver. It can be seen that during the loosening process, $r_{\rm sk}$ has different starting value, which gradually increases over time and eventually reaches the same maximum. On the contrary, during the tightening process, $r_{\rm sk}$ decreases from the same starting value, and end at a different radius. This is because $S(n)$ hosts more iSks than $S(n-1)$, which makes the $r_{\rm sk}$ of $S(n)$ larger. As a reference, radius variation of an iSk is also shown in Fig.~\ref{5} (a) by a dotted line, which decreases very slightly with the positive field. Notably, the fluctuation of $r_{\rm sk}$ at 11--14 ns is caused by the reconstruction of skyrmion bags induced by repulsive potential between a new iSk and original iSks. The static properties of skyrmion bags are shown in Fig.~\ref{5} (c), where the loosened and tightened states corresponding to Figs.~\ref{4} (a) (c) are marked by triangle and square symbols, respectively. Two parameters, $d_{\rm ss}$ and $d_{\rm sd}$ are introduced, with their definition shown schematically in the inset of Fig.~\ref{5} (d). $d_{\rm ss}$ is defined as the distance between two neighboring iSks, and $d_{\rm sd}$ is defined as the distance between an outermost iSk and the outer circular chiral domain wall of oSk. Overall, for loosened states, $d_{\rm ss}$ fluctuates up and down around 100 nm, while for tightened states, $d_{\rm ss}$ and $d_{\rm sd}$ both increase monotonically. It can be also seen that for a large topological charge such as $Q = 16$, the error value of $d_{\rm sd}$ becomes relatively large, but the error value of $d_{\rm ss}$ still remains in a small range. This can also suggest that the tightened state skyrmion bags are more stable than the loosened ones. In Fig.~\ref{5} (d), we give a prediction about how many iSks can be created in total on such a nanodisk. The prediction is carried out by introducing $r_{\rm max}$, defined as the distance from the outermost iSk to the center of the nanodisk, as schematically shown in the inset of Fig.~\ref{5} (d). The red dots are simulation results for $S(1)$ to $S(16)$, and the analytic curve is originated from extrema solutions suggested by Besicovitc lemma on Euclides geometry, given as \cite{bateman1951geometrical} 
	\begin{align}
		r_{\rm max}=d_{\rm ss}\cdot\sqrt{\frac{\sqrt{3}(Q+1)}{2\pi}}
		,\label{8}
	\end{align}
	where $d_{\rm ss}$ is taken as 100 nm according to Fig.~\ref{5} (c). Obviously, the theoretical curve is well matched with the simulation results for $S(1)$ to $S(16)$, so it is reasonable to believe that this curve can predict the subsequent trend. Considering that the distance from the nucleation area to the center is 384 nm, and the size of iSks is around 30 nm, we conservatively estimate that at least $S(40)$ can be created on this nanodisk.
	
	\begin{figure*}[t]
		\includegraphics[width=1\linewidth]{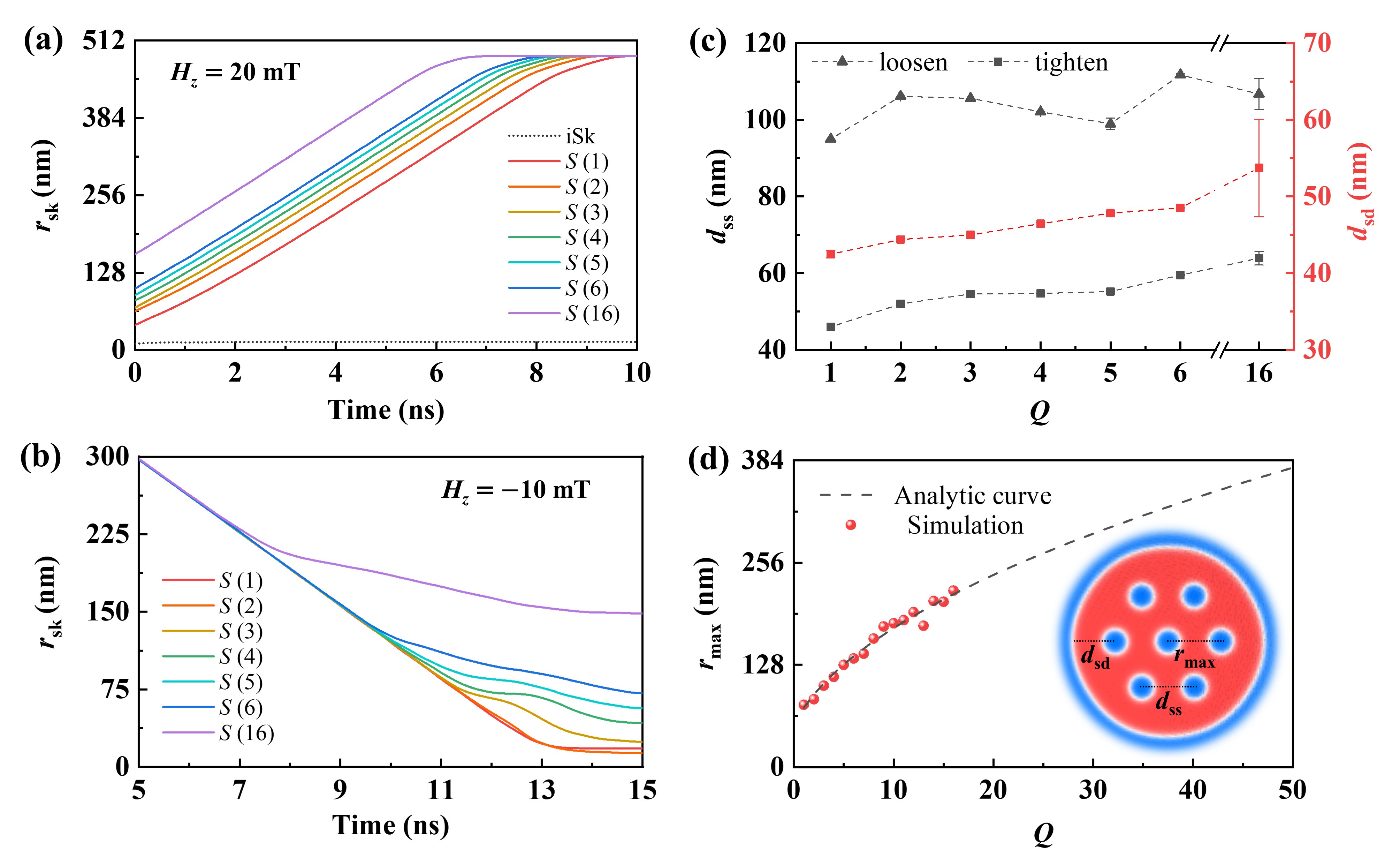}
		\caption{Dynamical variation of the oSk radius $r_{\rm sk}$ versus time, for $S(1)$ to $S(6)$, and $S(16)$, under (a) $H_z=20\ {\rm mT}$ and (b) $H_z=-10\ {\rm mT}$. Dotted line represents the variation of an iSk radius as a reference. (c) Variation of parameters of static properties, $d_{\rm ss}$ and $d_{\rm sd}$ versus $Q$, for both loosen and tighten skyrmion bags. The error bars give the variable ranges of corresponding parameters. (d) Variation and prediction of $r_{\rm max}$ versus $Q$. The red dot symbols represent simulation results, and the dashed line represents the analytic curve. Inset: schematic of the definitions of  $d_{\rm ss}$, $d_{\rm sd}$ and $r_{\rm max}$. $d_{\rm ss}$ is defined as the distance between two neighboring iSks; $d_{\rm sd}$ is defined as the distance between an outermost iSk and the outer circular chiral domain wall of oSk; $r_{\rm max}$ is defined as the distance from the outermost iSk to the center of the nanodisk.}\label{5}     
	\end{figure*}
	
	\section{Conclusion and Prospect}
	In conclusion, we proposed a method to create magnetic skyrmion bags with arbitrary topological charge in a 2D nanodisk. Firstly, we studied the field dependence and equilibrium states to determine the threshold value of external field for regulating the size of skyrmion bags. Then, as an example, we showed a complete cycle of creating $S(1)$, which includes three steps: loosening $S(0)$ by a positive field, nucleating an iSk by current pulses, and tightening $S(1)$ by a negative field. The nucleated iSk stability was mapped by a phase diagram with respect to current intensity and nanocontact diameter, and the mechanism was explained by introducing the system maximum torque. The tightened step was analyzed by a force diagram and Thiele equation, which suggest that the force from confining potential plays a key role in driving iSk towards the central area. Moreover, corresponding simulated L-TEM images were presented to facilitate the experimental investigation. Finally, we compared the static and dynamical properties of $S(1)$ to $S(16)$ to reveal the multiple inter-skyrmion interactions when skyrmion bags reconstructing. Based on analytical solution, we gave a prediction that at least $S(40)$ could be created by the proposed approach on this nanodik. 
	
	The present work primarily focuses on fundamental physics, and it acknowledges the presence of technical challenges that need to be addressed before experimental verification. For example, the skyrmion bags might get destabilized by the vertical field or even collapse at room temperature due to the reduction in energy barrier. To enhance the stability of room-temperature skyrmions, one experimental approach involves the construction of multilayers \cite{Anjan2017Tunable}. Another issue is about the realistic current required to create the iSks. On the one hand, supplemental materials related to cell discretization indicate that overcoming the local twist becomes more challenging in smaller cells, suggesting the possibility of higher actual threshold currents \cite{2018A}. On the other hand, the vertical current generates a swirling Oersted field in the plane, which was not considered in this work. It has been demonstrated that the Oersted field facilitates the creation of skyrmions \cite{2016Topological}, implying that the actual threshold currents may be reduced. We hope that our theoretical findings will stimulate future experimental works and encourage exploration of technological applications based on skyrmion bags.

	\begin{acknowledgments}
	The authors gratefully acknowledge the National Science Fund for Distinguished Young Scholars (52225312), the National Key Research and Development Program of China (2019YFE0121700), and the Key Research and Development Program of Zhejiang Province (2021C01033). L. Bo thanks the financial support by China Scholarship Council (No. 202206080023). This work was also supported by JSPS KAKENHI (20H00337) and JST CREST (JPMJCR20T1).
	\end{acknowledgments}
	
	\bibliographystyle{apsrev4-2}
	\bibliography{BoLan_SkyrmionBags}

\begin{thebibliography}{58}%
\makeatletter
\providecommand \@ifxundefined [1]{%
 \@ifx{#1\undefined}
}%
\providecommand \@ifnum [1]{%
 \ifnum #1\expandafter \@firstoftwo
 \else \expandafter \@secondoftwo
 \fi
}%
\providecommand \@ifx [1]{%
 \ifx #1\expandafter \@firstoftwo
 \else \expandafter \@secondoftwo
 \fi
}%
\providecommand \natexlab [1]{#1}%
\providecommand \enquote  [1]{``#1''}%
\providecommand \bibnamefont  [1]{#1}%
\providecommand \bibfnamefont [1]{#1}%
\providecommand \citenamefont [1]{#1}%
\providecommand \href@noop [0]{\@secondoftwo}%
\providecommand \href [0]{\begingroup \@sanitize@url \@href}%
\providecommand \@href[1]{\@@startlink{#1}\@@href}%
\providecommand \@@href[1]{\endgroup#1\@@endlink}%
\providecommand \@sanitize@url [0]{\catcode `\\12\catcode `\$12\catcode
  `\&12\catcode `\#12\catcode `\^12\catcode `\_12\catcode `\%12\relax}%
\providecommand \@@startlink[1]{}%
\providecommand \@@endlink[0]{}%
\providecommand \url  [0]{\begingroup\@sanitize@url \@url }%
\providecommand \@url [1]{\endgroup\@href {#1}{\urlprefix }}%
\providecommand \urlprefix  [0]{URL }%
\providecommand \Eprint [0]{\href }%
\providecommand \doibase [0]{https://doi.org/}%
\providecommand \selectlanguage [0]{\@gobble}%
\providecommand \bibinfo  [0]{\@secondoftwo}%
\providecommand \bibfield  [0]{\@secondoftwo}%
\providecommand \translation [1]{[#1]}%
\providecommand \BibitemOpen [0]{}%
\providecommand \bibitemStop [0]{}%
\providecommand \bibitemNoStop [0]{.\EOS\space}%
\providecommand \EOS [0]{\spacefactor3000\relax}%
\providecommand \BibitemShut  [1]{\csname bibitem#1\endcsname}%
\let\auto@bib@innerbib\@empty
\bibitem [{\citenamefont {Mochizuki}\ and\ \citenamefont
  {Seki}(2015)}]{mochizuki2015dynamical}%
  \BibitemOpen
  \bibfield  {author} {\bibinfo {author} {\bibfnamefont {M.}~\bibnamefont
  {Mochizuki}}\ and\ \bibinfo {author} {\bibfnamefont {S.}~\bibnamefont
  {Seki}},\ }\href@noop {} {\bibfield  {journal} {\bibinfo  {journal} {Journal
  of Physics: Condensed Matter}\ }\textbf {\bibinfo {volume} {27}},\ \bibinfo
  {pages} {503001} (\bibinfo {year} {2015})}\BibitemShut {NoStop}%
\bibitem [{\citenamefont {Zhang}\ \emph {et~al.}(2020)\citenamefont {Zhang},
  \citenamefont {Zhou}, \citenamefont {Song}, \citenamefont {Park},
  \citenamefont {Xia}, \citenamefont {Ezawa}, \citenamefont {Liu},
  \citenamefont {Zhao}, \citenamefont {Zhao},\ and\ \citenamefont
  {Woo}}]{zhang2020skyrmion}%
  \BibitemOpen
  \bibfield  {author} {\bibinfo {author} {\bibfnamefont {X.}~\bibnamefont
  {Zhang}}, \bibinfo {author} {\bibfnamefont {Y.}~\bibnamefont {Zhou}},
  \bibinfo {author} {\bibfnamefont {K.~M.}\ \bibnamefont {Song}}, \bibinfo
  {author} {\bibfnamefont {T.-E.}\ \bibnamefont {Park}}, \bibinfo {author}
  {\bibfnamefont {J.}~\bibnamefont {Xia}}, \bibinfo {author} {\bibfnamefont
  {M.}~\bibnamefont {Ezawa}}, \bibinfo {author} {\bibfnamefont
  {X.}~\bibnamefont {Liu}}, \bibinfo {author} {\bibfnamefont {W.}~\bibnamefont
  {Zhao}}, \bibinfo {author} {\bibfnamefont {G.}~\bibnamefont {Zhao}},\ and\
  \bibinfo {author} {\bibfnamefont {S.}~\bibnamefont {Woo}},\ }\href@noop {}
  {\bibfield  {journal} {\bibinfo  {journal} {Journal of Physics: Condensed
  Matter}\ }\textbf {\bibinfo {volume} {32}},\ \bibinfo {pages} {143001}
  (\bibinfo {year} {2020})}\BibitemShut {NoStop}%
\bibitem [{\citenamefont {Bo}\ \emph {et~al.}(2022{\natexlab{a}})\citenamefont
  {Bo}, \citenamefont {Hu}, \citenamefont {Zhao},\ and\ \citenamefont
  {Zhang}}]{bo2022micromagnetic}%
  \BibitemOpen
  \bibfield  {author} {\bibinfo {author} {\bibfnamefont {L.}~\bibnamefont
  {Bo}}, \bibinfo {author} {\bibfnamefont {C.}~\bibnamefont {Hu}}, \bibinfo
  {author} {\bibfnamefont {R.}~\bibnamefont {Zhao}},\ and\ \bibinfo {author}
  {\bibfnamefont {X.}~\bibnamefont {Zhang}},\ }\href@noop {} {\bibfield
  {journal} {\bibinfo  {journal} {Journal of Physics D: Applied Physics}\ }
  (\bibinfo {year} {2022}{\natexlab{a}})}\BibitemShut {NoStop}%
\bibitem [{\citenamefont {Yu}\ \emph {et~al.}(2010)\citenamefont {Yu},
  \citenamefont {Onose}, \citenamefont {Kanazawa}, \citenamefont {Park},
  \citenamefont {Han}, \citenamefont {Matsui}, \citenamefont {Nagaosa},\ and\
  \citenamefont {Tokura}}]{yu2010real}%
  \BibitemOpen
  \bibfield  {author} {\bibinfo {author} {\bibfnamefont {X.}~\bibnamefont
  {Yu}}, \bibinfo {author} {\bibfnamefont {Y.}~\bibnamefont {Onose}}, \bibinfo
  {author} {\bibfnamefont {N.}~\bibnamefont {Kanazawa}}, \bibinfo {author}
  {\bibfnamefont {J.~H.}\ \bibnamefont {Park}}, \bibinfo {author}
  {\bibfnamefont {J.}~\bibnamefont {Han}}, \bibinfo {author} {\bibfnamefont
  {Y.}~\bibnamefont {Matsui}}, \bibinfo {author} {\bibfnamefont
  {N.}~\bibnamefont {Nagaosa}},\ and\ \bibinfo {author} {\bibfnamefont
  {Y.}~\bibnamefont {Tokura}},\ }\href@noop {} {\bibfield  {journal} {\bibinfo
  {journal} {Nature}\ }\textbf {\bibinfo {volume} {465}},\ \bibinfo {pages}
  {901} (\bibinfo {year} {2010})}\BibitemShut {NoStop}%
\bibitem [{\citenamefont {G{\"o}bel}\ \emph {et~al.}(2021)\citenamefont
  {G{\"o}bel}, \citenamefont {Mertig},\ and\ \citenamefont
  {Tretiakov}}]{gobel2021beyond}%
  \BibitemOpen
  \bibfield  {author} {\bibinfo {author} {\bibfnamefont {B.}~\bibnamefont
  {G{\"o}bel}}, \bibinfo {author} {\bibfnamefont {I.}~\bibnamefont {Mertig}},\
  and\ \bibinfo {author} {\bibfnamefont {O.~A.}\ \bibnamefont {Tretiakov}},\
  }\href@noop {} {\bibfield  {journal} {\bibinfo  {journal} {Physics Reports}\
  }\textbf {\bibinfo {volume} {895}},\ \bibinfo {pages} {1} (\bibinfo {year}
  {2021})}\BibitemShut {NoStop}%
\bibitem [{\citenamefont {Foster}\ \emph {et~al.}(2019)\citenamefont {Foster},
  \citenamefont {Kind}, \citenamefont {Ackerman}, \citenamefont {Tai},
  \citenamefont {Dennis},\ and\ \citenamefont {Smalyukh}}]{foster2019two}%
  \BibitemOpen
  \bibfield  {author} {\bibinfo {author} {\bibfnamefont {D.}~\bibnamefont
  {Foster}}, \bibinfo {author} {\bibfnamefont {C.}~\bibnamefont {Kind}},
  \bibinfo {author} {\bibfnamefont {P.~J.}\ \bibnamefont {Ackerman}}, \bibinfo
  {author} {\bibfnamefont {J.-S.~B.}\ \bibnamefont {Tai}}, \bibinfo {author}
  {\bibfnamefont {M.~R.}\ \bibnamefont {Dennis}},\ and\ \bibinfo {author}
  {\bibfnamefont {I.~I.}\ \bibnamefont {Smalyukh}},\ }\href@noop {} {\bibfield
  {journal} {\bibinfo  {journal} {Nature Physics}\ }\textbf {\bibinfo {volume}
  {15}},\ \bibinfo {pages} {655} (\bibinfo {year} {2019})}\BibitemShut
  {NoStop}%
\bibitem [{\citenamefont {Rybakov}\ and\ \citenamefont
  {Kiselev}(2019)}]{rybakov2019chiral}%
  \BibitemOpen
  \bibfield  {author} {\bibinfo {author} {\bibfnamefont {F.~N.}\ \bibnamefont
  {Rybakov}}\ and\ \bibinfo {author} {\bibfnamefont {N.~S.}\ \bibnamefont
  {Kiselev}},\ }\href@noop {} {\bibfield  {journal} {\bibinfo  {journal}
  {Physical Review B}\ }\textbf {\bibinfo {volume} {99}},\ \bibinfo {pages}
  {064437} (\bibinfo {year} {2019})}\BibitemShut {NoStop}%
\bibitem [{\citenamefont {Kind}\ \emph {et~al.}(2020)\citenamefont {Kind},
  \citenamefont {Friedemann},\ and\ \citenamefont {Read}}]{kind2020existence}%
  \BibitemOpen
  \bibfield  {author} {\bibinfo {author} {\bibfnamefont {C.}~\bibnamefont
  {Kind}}, \bibinfo {author} {\bibfnamefont {S.}~\bibnamefont {Friedemann}},\
  and\ \bibinfo {author} {\bibfnamefont {D.}~\bibnamefont {Read}},\ }\href@noop
  {} {\bibfield  {journal} {\bibinfo  {journal} {Applied Physics Letters}\
  }\textbf {\bibinfo {volume} {116}},\ \bibinfo {pages} {022413} (\bibinfo
  {year} {2020})}\BibitemShut {NoStop}%
\bibitem [{\citenamefont {Zeng}\ \emph
  {et~al.}(2022{\natexlab{a}})\citenamefont {Zeng}, \citenamefont {Song},
  \citenamefont {Wang},\ and\ \citenamefont {Liu}}]{zeng2022spin}%
  \BibitemOpen
  \bibfield  {author} {\bibinfo {author} {\bibfnamefont {Z.}~\bibnamefont
  {Zeng}}, \bibinfo {author} {\bibfnamefont {C.}~\bibnamefont {Song}}, \bibinfo
  {author} {\bibfnamefont {J.}~\bibnamefont {Wang}},\ and\ \bibinfo {author}
  {\bibfnamefont {Q.}~\bibnamefont {Liu}},\ }\href@noop {} {\bibfield
  {journal} {\bibinfo  {journal} {Journal of Physics D: Applied Physics}\
  }\textbf {\bibinfo {volume} {55}},\ \bibinfo {pages} {185001} (\bibinfo
  {year} {2022}{\natexlab{a}})}\BibitemShut {NoStop}%
\bibitem [{\citenamefont {Zeng}\ \emph {et~al.}(2020)\citenamefont {Zeng},
  \citenamefont {Zhang}, \citenamefont {Jin}, \citenamefont {Wang},
  \citenamefont {Song}, \citenamefont {Ma}, \citenamefont {Liu},\ and\
  \citenamefont {Wang}}]{zeng2020dynamics}%
  \BibitemOpen
  \bibfield  {author} {\bibinfo {author} {\bibfnamefont {Z.}~\bibnamefont
  {Zeng}}, \bibinfo {author} {\bibfnamefont {C.}~\bibnamefont {Zhang}},
  \bibinfo {author} {\bibfnamefont {C.}~\bibnamefont {Jin}}, \bibinfo {author}
  {\bibfnamefont {J.}~\bibnamefont {Wang}}, \bibinfo {author} {\bibfnamefont
  {C.}~\bibnamefont {Song}}, \bibinfo {author} {\bibfnamefont {Y.}~\bibnamefont
  {Ma}}, \bibinfo {author} {\bibfnamefont {Q.}~\bibnamefont {Liu}},\ and\
  \bibinfo {author} {\bibfnamefont {J.}~\bibnamefont {Wang}},\ }\href@noop {}
  {\bibfield  {journal} {\bibinfo  {journal} {Applied Physics Letters}\
  }\textbf {\bibinfo {volume} {117}},\ \bibinfo {pages} {172404} (\bibinfo
  {year} {2020})}\BibitemShut {NoStop}%
\bibitem [{\citenamefont {Kind}\ and\ \citenamefont
  {Foster}(2021)}]{kind2021magnetic}%
  \BibitemOpen
  \bibfield  {author} {\bibinfo {author} {\bibfnamefont {C.}~\bibnamefont
  {Kind}}\ and\ \bibinfo {author} {\bibfnamefont {D.}~\bibnamefont {Foster}},\
  }\href@noop {} {\bibfield  {journal} {\bibinfo  {journal} {Physical Review
  B}\ }\textbf {\bibinfo {volume} {103}},\ \bibinfo {pages} {L100413} (\bibinfo
  {year} {2021})}\BibitemShut {NoStop}%
\bibitem [{\citenamefont {Zeng}\ \emph
  {et~al.}(2022{\natexlab{b}})\citenamefont {Zeng}, \citenamefont {Mehmood},
  \citenamefont {Ma}, \citenamefont {Wang}, \citenamefont {Wang},\ and\
  \citenamefont {Liu}}]{zeng2022skyrmion}%
  \BibitemOpen
  \bibfield  {author} {\bibinfo {author} {\bibfnamefont {Z.}~\bibnamefont
  {Zeng}}, \bibinfo {author} {\bibfnamefont {N.}~\bibnamefont {Mehmood}},
  \bibinfo {author} {\bibfnamefont {Y.}~\bibnamefont {Ma}}, \bibinfo {author}
  {\bibfnamefont {J.}~\bibnamefont {Wang}}, \bibinfo {author} {\bibfnamefont
  {J.}~\bibnamefont {Wang}},\ and\ \bibinfo {author} {\bibfnamefont
  {Q.}~\bibnamefont {Liu}},\ }\href@noop {} {\bibfield  {journal} {\bibinfo
  {journal} {arXiv preprint arXiv:2204.04439}\ } (\bibinfo {year}
  {2022}{\natexlab{b}})}\BibitemShut {NoStop}%
\bibitem [{\citenamefont
  {Dzyaloshinsky}(1958)}]{dzyaloshinsky1958thermodynamic}%
  \BibitemOpen
  \bibfield  {author} {\bibinfo {author} {\bibfnamefont {I.}~\bibnamefont
  {Dzyaloshinsky}},\ }\href@noop {} {\bibfield  {journal} {\bibinfo  {journal}
  {Journal of physics and chemistry of solids}\ }\textbf {\bibinfo {volume}
  {4}},\ \bibinfo {pages} {241} (\bibinfo {year} {1958})}\BibitemShut {NoStop}%
\bibitem [{\citenamefont {Moriya}(1960)}]{moriya1960anisotropic}%
  \BibitemOpen
  \bibfield  {author} {\bibinfo {author} {\bibfnamefont {T.}~\bibnamefont
  {Moriya}},\ }\href@noop {} {\bibfield  {journal} {\bibinfo  {journal}
  {Physical review}\ }\textbf {\bibinfo {volume} {120}},\ \bibinfo {pages} {91}
  (\bibinfo {year} {1960})}\BibitemShut {NoStop}%
\bibitem [{\citenamefont {Du}\ \emph {et~al.}(2015)\citenamefont {Du},
  \citenamefont {Che}, \citenamefont {Kong}, \citenamefont {Zhao},
  \citenamefont {Jin}, \citenamefont {Wang}, \citenamefont {Yang},
  \citenamefont {Ning}, \citenamefont {Li}, \citenamefont {Jin} \emph
  {et~al.}}]{du2015edge}%
  \BibitemOpen
  \bibfield  {author} {\bibinfo {author} {\bibfnamefont {H.}~\bibnamefont
  {Du}}, \bibinfo {author} {\bibfnamefont {R.}~\bibnamefont {Che}}, \bibinfo
  {author} {\bibfnamefont {L.}~\bibnamefont {Kong}}, \bibinfo {author}
  {\bibfnamefont {X.}~\bibnamefont {Zhao}}, \bibinfo {author} {\bibfnamefont
  {C.}~\bibnamefont {Jin}}, \bibinfo {author} {\bibfnamefont {C.}~\bibnamefont
  {Wang}}, \bibinfo {author} {\bibfnamefont {J.}~\bibnamefont {Yang}}, \bibinfo
  {author} {\bibfnamefont {W.}~\bibnamefont {Ning}}, \bibinfo {author}
  {\bibfnamefont {R.}~\bibnamefont {Li}}, \bibinfo {author} {\bibfnamefont
  {C.}~\bibnamefont {Jin}}, \emph {et~al.},\ }\href@noop {} {\bibfield
  {journal} {\bibinfo  {journal} {Nature communications}\ }\textbf {\bibinfo
  {volume} {6}},\ \bibinfo {pages} {1} (\bibinfo {year} {2015})}\BibitemShut
  {NoStop}%
\bibitem [{\citenamefont {Leonov}(2022)}]{leonov2022skyrmion}%
  \BibitemOpen
  \bibfield  {author} {\bibinfo {author} {\bibfnamefont {A.~O.}\ \bibnamefont
  {Leonov}},\ }\href@noop {} {\bibfield  {journal} {\bibinfo  {journal}
  {Physical Review B}\ }\textbf {\bibinfo {volume} {105}},\ \bibinfo {pages}
  {094404} (\bibinfo {year} {2022})}\BibitemShut {NoStop}%
\bibitem [{\citenamefont {Zhang}\ \emph
  {et~al.}(2017{\natexlab{a}})\citenamefont {Zhang}, \citenamefont {Xia},
  \citenamefont {Zhou}, \citenamefont {Liu}, \citenamefont {Zhang},\ and\
  \citenamefont {Ezawa}}]{zhang2017skyrmion}%
  \BibitemOpen
  \bibfield  {author} {\bibinfo {author} {\bibfnamefont {X.}~\bibnamefont
  {Zhang}}, \bibinfo {author} {\bibfnamefont {J.}~\bibnamefont {Xia}}, \bibinfo
  {author} {\bibfnamefont {Y.}~\bibnamefont {Zhou}}, \bibinfo {author}
  {\bibfnamefont {X.}~\bibnamefont {Liu}}, \bibinfo {author} {\bibfnamefont
  {H.}~\bibnamefont {Zhang}},\ and\ \bibinfo {author} {\bibfnamefont
  {M.}~\bibnamefont {Ezawa}},\ }\href@noop {} {\bibfield  {journal} {\bibinfo
  {journal} {Nature communications}\ }\textbf {\bibinfo {volume} {8}},\
  \bibinfo {pages} {1} (\bibinfo {year} {2017}{\natexlab{a}})}\BibitemShut
  {NoStop}%
\bibitem [{\citenamefont {Jiang}\ \emph {et~al.}(2018)\citenamefont {Jiang},
  \citenamefont {Xia}, \citenamefont {Zhang}, \citenamefont {Song},
  \citenamefont {Ma}, \citenamefont {Fangohr}, \citenamefont {Zhao},
  \citenamefont {Liu}, \citenamefont {Zhao},\ and\ \citenamefont
  {Zhou}}]{jiang2018dynamics}%
  \BibitemOpen
  \bibfield  {author} {\bibinfo {author} {\bibfnamefont {W.}~\bibnamefont
  {Jiang}}, \bibinfo {author} {\bibfnamefont {J.}~\bibnamefont {Xia}}, \bibinfo
  {author} {\bibfnamefont {X.}~\bibnamefont {Zhang}}, \bibinfo {author}
  {\bibfnamefont {Y.}~\bibnamefont {Song}}, \bibinfo {author} {\bibfnamefont
  {C.}~\bibnamefont {Ma}}, \bibinfo {author} {\bibfnamefont {H.}~\bibnamefont
  {Fangohr}}, \bibinfo {author} {\bibfnamefont {G.}~\bibnamefont {Zhao}},
  \bibinfo {author} {\bibfnamefont {X.}~\bibnamefont {Liu}}, \bibinfo {author}
  {\bibfnamefont {W.}~\bibnamefont {Zhao}},\ and\ \bibinfo {author}
  {\bibfnamefont {Y.}~\bibnamefont {Zhou}},\ }\href@noop {} {\bibfield
  {journal} {\bibinfo  {journal} {IEEE Magnetics Letters}\ }\textbf {\bibinfo
  {volume} {9}},\ \bibinfo {pages} {1} (\bibinfo {year} {2018})}\BibitemShut
  {NoStop}%
\bibitem [{\citenamefont {Leonov}\ and\ \citenamefont
  {Pappas}(2019)}]{leonov2019skyrmion}%
  \BibitemOpen
  \bibfield  {author} {\bibinfo {author} {\bibfnamefont {A.}~\bibnamefont
  {Leonov}}\ and\ \bibinfo {author} {\bibfnamefont {C.}~\bibnamefont
  {Pappas}},\ }\href@noop {} {\bibfield  {journal} {\bibinfo  {journal}
  {Physical Review B}\ }\textbf {\bibinfo {volume} {99}},\ \bibinfo {pages}
  {144410} (\bibinfo {year} {2019})}\BibitemShut {NoStop}%
\bibitem [{\citenamefont {Zheng}\ \emph {et~al.}(2021)\citenamefont {Zheng},
  \citenamefont {Rybakov}, \citenamefont {Kiselev}, \citenamefont {Song},
  \citenamefont {Kov{\'a}cs}, \citenamefont {Du}, \citenamefont {Bl{\"u}gel},\
  and\ \citenamefont {Dunin-Borkowski}}]{zheng2021magnetic}%
  \BibitemOpen
  \bibfield  {author} {\bibinfo {author} {\bibfnamefont {F.}~\bibnamefont
  {Zheng}}, \bibinfo {author} {\bibfnamefont {F.~N.}\ \bibnamefont {Rybakov}},
  \bibinfo {author} {\bibfnamefont {N.~S.}\ \bibnamefont {Kiselev}}, \bibinfo
  {author} {\bibfnamefont {D.}~\bibnamefont {Song}}, \bibinfo {author}
  {\bibfnamefont {A.}~\bibnamefont {Kov{\'a}cs}}, \bibinfo {author}
  {\bibfnamefont {H.}~\bibnamefont {Du}}, \bibinfo {author} {\bibfnamefont
  {S.}~\bibnamefont {Bl{\"u}gel}},\ and\ \bibinfo {author} {\bibfnamefont
  {R.~E.}\ \bibnamefont {Dunin-Borkowski}},\ }\href@noop {} {\bibfield
  {journal} {\bibinfo  {journal} {Nature communications}\ }\textbf {\bibinfo
  {volume} {12}},\ \bibinfo {pages} {1} (\bibinfo {year} {2021})}\BibitemShut
  {NoStop}%
\bibitem [{\citenamefont {Tang}\ \emph {et~al.}(2021)\citenamefont {Tang},
  \citenamefont {Wu}, \citenamefont {Wang}, \citenamefont {Kong}, \citenamefont
  {Lv}, \citenamefont {Wei}, \citenamefont {Zang}, \citenamefont {Tian},\ and\
  \citenamefont {Du}}]{tang2021magnetic}%
  \BibitemOpen
  \bibfield  {author} {\bibinfo {author} {\bibfnamefont {J.}~\bibnamefont
  {Tang}}, \bibinfo {author} {\bibfnamefont {Y.}~\bibnamefont {Wu}}, \bibinfo
  {author} {\bibfnamefont {W.}~\bibnamefont {Wang}}, \bibinfo {author}
  {\bibfnamefont {L.}~\bibnamefont {Kong}}, \bibinfo {author} {\bibfnamefont
  {B.}~\bibnamefont {Lv}}, \bibinfo {author} {\bibfnamefont {W.}~\bibnamefont
  {Wei}}, \bibinfo {author} {\bibfnamefont {J.}~\bibnamefont {Zang}}, \bibinfo
  {author} {\bibfnamefont {M.}~\bibnamefont {Tian}},\ and\ \bibinfo {author}
  {\bibfnamefont {H.}~\bibnamefont {Du}},\ }\href@noop {} {\bibfield  {journal}
  {\bibinfo  {journal} {Nature Nanotechnology}\ }\textbf {\bibinfo {volume}
  {16}},\ \bibinfo {pages} {1086} (\bibinfo {year} {2021})}\BibitemShut
  {NoStop}%
\bibitem [{\citenamefont {Tchoe}\ and\ \citenamefont
  {Han}(2012)}]{tchoe2012skyrmion}%
  \BibitemOpen
  \bibfield  {author} {\bibinfo {author} {\bibfnamefont {Y.}~\bibnamefont
  {Tchoe}}\ and\ \bibinfo {author} {\bibfnamefont {J.~H.}\ \bibnamefont
  {Han}},\ }\href@noop {} {\bibfield  {journal} {\bibinfo  {journal} {Physical
  Review B}\ }\textbf {\bibinfo {volume} {85}},\ \bibinfo {pages} {174416}
  (\bibinfo {year} {2012})}\BibitemShut {NoStop}%
\bibitem [{\citenamefont {Iwasaki}\ \emph {et~al.}(2013)\citenamefont
  {Iwasaki}, \citenamefont {Mochizuki},\ and\ \citenamefont
  {Nagaosa}}]{iwasaki2013current}%
  \BibitemOpen
  \bibfield  {author} {\bibinfo {author} {\bibfnamefont {J.}~\bibnamefont
  {Iwasaki}}, \bibinfo {author} {\bibfnamefont {M.}~\bibnamefont {Mochizuki}},\
  and\ \bibinfo {author} {\bibfnamefont {N.}~\bibnamefont {Nagaosa}},\
  }\href@noop {} {\bibfield  {journal} {\bibinfo  {journal} {Nature
  nanotechnology}\ }\textbf {\bibinfo {volume} {8}},\ \bibinfo {pages} {742}
  (\bibinfo {year} {2013})}\BibitemShut {NoStop}%
\bibitem [{\citenamefont {Jiang}\ \emph {et~al.}(2015)\citenamefont {Jiang},
  \citenamefont {Upadhyaya}, \citenamefont {Zhang}, \citenamefont {Yu},
  \citenamefont {Jungfleisch}, \citenamefont {Fradin}, \citenamefont {Pearson},
  \citenamefont {Tserkovnyak}, \citenamefont {Wang}, \citenamefont {Heinonen}
  \emph {et~al.}}]{jiang2015blowing}%
  \BibitemOpen
  \bibfield  {author} {\bibinfo {author} {\bibfnamefont {W.}~\bibnamefont
  {Jiang}}, \bibinfo {author} {\bibfnamefont {P.}~\bibnamefont {Upadhyaya}},
  \bibinfo {author} {\bibfnamefont {W.}~\bibnamefont {Zhang}}, \bibinfo
  {author} {\bibfnamefont {G.}~\bibnamefont {Yu}}, \bibinfo {author}
  {\bibfnamefont {M.~B.}\ \bibnamefont {Jungfleisch}}, \bibinfo {author}
  {\bibfnamefont {F.~Y.}\ \bibnamefont {Fradin}}, \bibinfo {author}
  {\bibfnamefont {J.~E.}\ \bibnamefont {Pearson}}, \bibinfo {author}
  {\bibfnamefont {Y.}~\bibnamefont {Tserkovnyak}}, \bibinfo {author}
  {\bibfnamefont {K.~L.}\ \bibnamefont {Wang}}, \bibinfo {author}
  {\bibfnamefont {O.}~\bibnamefont {Heinonen}}, \emph {et~al.},\ }\href@noop {}
  {\bibfield  {journal} {\bibinfo  {journal} {Science}\ }\textbf {\bibinfo
  {volume} {349}},\ \bibinfo {pages} {283} (\bibinfo {year}
  {2015})}\BibitemShut {NoStop}%
\bibitem [{\citenamefont {Lin}(2016)}]{lin2016edge}%
  \BibitemOpen
  \bibfield  {author} {\bibinfo {author} {\bibfnamefont {S.-Z.}\ \bibnamefont
  {Lin}},\ }\href@noop {} {\bibfield  {journal} {\bibinfo  {journal} {Physical
  Review B}\ }\textbf {\bibinfo {volume} {94}},\ \bibinfo {pages} {020402}
  (\bibinfo {year} {2016})}\BibitemShut {NoStop}%
\bibitem [{\citenamefont {Heinonen}\ \emph {et~al.}(2016)\citenamefont
  {Heinonen}, \citenamefont {Jiang}, \citenamefont {Somaily}, \citenamefont
  {Te~Velthuis},\ and\ \citenamefont {Hoffmann}}]{heinonen2016generation}%
  \BibitemOpen
  \bibfield  {author} {\bibinfo {author} {\bibfnamefont {O.}~\bibnamefont
  {Heinonen}}, \bibinfo {author} {\bibfnamefont {W.}~\bibnamefont {Jiang}},
  \bibinfo {author} {\bibfnamefont {H.}~\bibnamefont {Somaily}}, \bibinfo
  {author} {\bibfnamefont {S.~G.}\ \bibnamefont {Te~Velthuis}},\ and\ \bibinfo
  {author} {\bibfnamefont {A.}~\bibnamefont {Hoffmann}},\ }\href@noop {}
  {\bibfield  {journal} {\bibinfo  {journal} {Physical Review B}\ }\textbf
  {\bibinfo {volume} {93}},\ \bibinfo {pages} {094407} (\bibinfo {year}
  {2016})}\BibitemShut {NoStop}%
\bibitem [{\citenamefont {Wang}\ \emph {et~al.}(2022)\citenamefont {Wang},
  \citenamefont {Song}, \citenamefont {Wei}, \citenamefont {Nan}, \citenamefont
  {Zhang}, \citenamefont {Ge}, \citenamefont {Tian}, \citenamefont {Zang},\
  and\ \citenamefont {Du}}]{wang2022electrical}%
  \BibitemOpen
  \bibfield  {author} {\bibinfo {author} {\bibfnamefont {W.}~\bibnamefont
  {Wang}}, \bibinfo {author} {\bibfnamefont {D.}~\bibnamefont {Song}}, \bibinfo
  {author} {\bibfnamefont {W.}~\bibnamefont {Wei}}, \bibinfo {author}
  {\bibfnamefont {P.}~\bibnamefont {Nan}}, \bibinfo {author} {\bibfnamefont
  {S.}~\bibnamefont {Zhang}}, \bibinfo {author} {\bibfnamefont
  {B.}~\bibnamefont {Ge}}, \bibinfo {author} {\bibfnamefont {M.}~\bibnamefont
  {Tian}}, \bibinfo {author} {\bibfnamefont {J.}~\bibnamefont {Zang}},\ and\
  \bibinfo {author} {\bibfnamefont {H.}~\bibnamefont {Du}},\ }\href@noop {}
  {\bibfield  {journal} {\bibinfo  {journal} {Nature communications}\ }\textbf
  {\bibinfo {volume} {13}},\ \bibinfo {pages} {1} (\bibinfo {year}
  {2022})}\BibitemShut {NoStop}%
\bibitem [{\citenamefont {Nagaosa}\ and\ \citenamefont
  {Tokura}(2013)}]{nagaosa2013topological}%
  \BibitemOpen
  \bibfield  {author} {\bibinfo {author} {\bibfnamefont {N.}~\bibnamefont
  {Nagaosa}}\ and\ \bibinfo {author} {\bibfnamefont {Y.}~\bibnamefont
  {Tokura}},\ }\href@noop {} {\bibfield  {journal} {\bibinfo  {journal} {Nature
  nanotechnology}\ }\textbf {\bibinfo {volume} {8}},\ \bibinfo {pages} {899}
  (\bibinfo {year} {2013})}\BibitemShut {NoStop}%
\bibitem [{\citenamefont {Bogdanov}\ and\ \citenamefont
  {Hubert}(1999)}]{bogdanov1999stability}%
  \BibitemOpen
  \bibfield  {author} {\bibinfo {author} {\bibfnamefont {A.}~\bibnamefont
  {Bogdanov}}\ and\ \bibinfo {author} {\bibfnamefont {A.}~\bibnamefont
  {Hubert}},\ }\href@noop {} {\bibfield  {journal} {\bibinfo  {journal}
  {Journal of magnetism and magnetic materials}\ }\textbf {\bibinfo {volume}
  {195}},\ \bibinfo {pages} {182} (\bibinfo {year} {1999})}\BibitemShut
  {NoStop}%
\bibitem [{\citenamefont {Zhang}\ \emph {et~al.}(2016)\citenamefont {Zhang},
  \citenamefont {Xia}, \citenamefont {Zhou}, \citenamefont {Wang},
  \citenamefont {Liu}, \citenamefont {Zhao},\ and\ \citenamefont
  {Ezawa}}]{zhang2016control}%
  \BibitemOpen
  \bibfield  {author} {\bibinfo {author} {\bibfnamefont {X.}~\bibnamefont
  {Zhang}}, \bibinfo {author} {\bibfnamefont {J.}~\bibnamefont {Xia}}, \bibinfo
  {author} {\bibfnamefont {Y.}~\bibnamefont {Zhou}}, \bibinfo {author}
  {\bibfnamefont {D.}~\bibnamefont {Wang}}, \bibinfo {author} {\bibfnamefont
  {X.}~\bibnamefont {Liu}}, \bibinfo {author} {\bibfnamefont {W.}~\bibnamefont
  {Zhao}},\ and\ \bibinfo {author} {\bibfnamefont {M.}~\bibnamefont {Ezawa}},\
  }\href@noop {} {\bibfield  {journal} {\bibinfo  {journal} {Physical Review
  B}\ }\textbf {\bibinfo {volume} {94}},\ \bibinfo {pages} {094420} (\bibinfo
  {year} {2016})}\BibitemShut {NoStop}%
\bibitem [{\citenamefont {D{\"u}rrenfeld}\ \emph {et~al.}(2017)\citenamefont
  {D{\"u}rrenfeld}, \citenamefont {Xu}, \citenamefont {{\AA}kerman},\ and\
  \citenamefont {Zhou}}]{durrenfeld2017controlled}%
  \BibitemOpen
  \bibfield  {author} {\bibinfo {author} {\bibfnamefont {P.}~\bibnamefont
  {D{\"u}rrenfeld}}, \bibinfo {author} {\bibfnamefont {Y.}~\bibnamefont {Xu}},
  \bibinfo {author} {\bibfnamefont {J.}~\bibnamefont {{\AA}kerman}},\ and\
  \bibinfo {author} {\bibfnamefont {Y.}~\bibnamefont {Zhou}},\ }\href@noop {}
  {\bibfield  {journal} {\bibinfo  {journal} {Physical Review B}\ }\textbf
  {\bibinfo {volume} {96}},\ \bibinfo {pages} {054430} (\bibinfo {year}
  {2017})}\BibitemShut {NoStop}%
\bibitem [{\citenamefont {Liu}\ \emph {et~al.}(2015)\citenamefont {Liu},
  \citenamefont {Du}, \citenamefont {Jia},\ and\ \citenamefont
  {Du}}]{liu2015switching}%
  \BibitemOpen
  \bibfield  {author} {\bibinfo {author} {\bibfnamefont {Y.}~\bibnamefont
  {Liu}}, \bibinfo {author} {\bibfnamefont {H.}~\bibnamefont {Du}}, \bibinfo
  {author} {\bibfnamefont {M.}~\bibnamefont {Jia}},\ and\ \bibinfo {author}
  {\bibfnamefont {A.}~\bibnamefont {Du}},\ }\href@noop {} {\bibfield  {journal}
  {\bibinfo  {journal} {Physical Review B}\ }\textbf {\bibinfo {volume} {91}},\
  \bibinfo {pages} {094425} (\bibinfo {year} {2015})}\BibitemShut {NoStop}%
\bibitem [{\citenamefont {Slonczewski}(1996)}]{slonczewski1996current}%
  \BibitemOpen
  \bibfield  {author} {\bibinfo {author} {\bibfnamefont {J.~C.}\ \bibnamefont
  {Slonczewski}},\ }\href@noop {} {\bibfield  {journal} {\bibinfo  {journal}
  {Journal of Magnetism and Magnetic Materials}\ }\textbf {\bibinfo {volume}
  {159}},\ \bibinfo {pages} {L1} (\bibinfo {year} {1996})}\BibitemShut
  {NoStop}%
\bibitem [{\citenamefont {Vansteenkiste}\ \emph {et~al.}(2014)\citenamefont
  {Vansteenkiste}, \citenamefont {Leliaert}, \citenamefont {Dvornik},
  \citenamefont {Helsen}, \citenamefont {Garcia-Sanchez},\ and\ \citenamefont
  {Van~Waeyenberge}}]{vansteenkiste2014design}%
  \BibitemOpen
  \bibfield  {author} {\bibinfo {author} {\bibfnamefont {A.}~\bibnamefont
  {Vansteenkiste}}, \bibinfo {author} {\bibfnamefont {J.}~\bibnamefont
  {Leliaert}}, \bibinfo {author} {\bibfnamefont {M.}~\bibnamefont {Dvornik}},
  \bibinfo {author} {\bibfnamefont {M.}~\bibnamefont {Helsen}}, \bibinfo
  {author} {\bibfnamefont {F.}~\bibnamefont {Garcia-Sanchez}},\ and\ \bibinfo
  {author} {\bibfnamefont {B.}~\bibnamefont {Van~Waeyenberge}},\ }\href@noop {}
  {\bibfield  {journal} {\bibinfo  {journal} {AIP advances}\ }\textbf {\bibinfo
  {volume} {4}},\ \bibinfo {pages} {107133} (\bibinfo {year}
  {2014})}\BibitemShut {NoStop}%
\bibitem [{\citenamefont {Abo}\ \emph {et~al.}(2013)\citenamefont {Abo},
  \citenamefont {Hong}, \citenamefont {Park}, \citenamefont {Lee},
  \citenamefont {Lee},\ and\ \citenamefont {Choi}}]{abo2013definition}%
  \BibitemOpen
  \bibfield  {author} {\bibinfo {author} {\bibfnamefont {G.~S.}\ \bibnamefont
  {Abo}}, \bibinfo {author} {\bibfnamefont {Y.-K.}\ \bibnamefont {Hong}},
  \bibinfo {author} {\bibfnamefont {J.}~\bibnamefont {Park}}, \bibinfo {author}
  {\bibfnamefont {J.}~\bibnamefont {Lee}}, \bibinfo {author} {\bibfnamefont
  {W.}~\bibnamefont {Lee}},\ and\ \bibinfo {author} {\bibfnamefont {B.-C.}\
  \bibnamefont {Choi}},\ }\href@noop {} {\bibfield  {journal} {\bibinfo
  {journal} {IEEE Transactions on Magnetics}\ }\textbf {\bibinfo {volume}
  {49}},\ \bibinfo {pages} {4937} (\bibinfo {year} {2013})}\BibitemShut
  {NoStop}%
\bibitem [{\citenamefont {Sampaio}\ \emph {et~al.}(2013)\citenamefont
  {Sampaio}, \citenamefont {Cros}, \citenamefont {Rohart}, \citenamefont
  {Thiaville},\ and\ \citenamefont {Fert}}]{sampaio2013nucleation}%
  \BibitemOpen
  \bibfield  {author} {\bibinfo {author} {\bibfnamefont {J.}~\bibnamefont
  {Sampaio}}, \bibinfo {author} {\bibfnamefont {V.}~\bibnamefont {Cros}},
  \bibinfo {author} {\bibfnamefont {S.}~\bibnamefont {Rohart}}, \bibinfo
  {author} {\bibfnamefont {A.}~\bibnamefont {Thiaville}},\ and\ \bibinfo
  {author} {\bibfnamefont {A.}~\bibnamefont {Fert}},\ }\href@noop {} {\bibfield
   {journal} {\bibinfo  {journal} {Nature nanotechnology}\ }\textbf {\bibinfo
  {volume} {8}},\ \bibinfo {pages} {839} (\bibinfo {year} {2013})}\BibitemShut
  {NoStop}%
\bibitem [{\citenamefont {Metaxas}\ \emph {et~al.}(2007)\citenamefont
  {Metaxas}, \citenamefont {Jamet}, \citenamefont {Mougin}, \citenamefont
  {Cormier}, \citenamefont {Ferr{\'e}}, \citenamefont {Baltz}, \citenamefont
  {Rodmacq}, \citenamefont {Dieny},\ and\ \citenamefont
  {Stamps}}]{metaxas2007creep}%
  \BibitemOpen
  \bibfield  {author} {\bibinfo {author} {\bibfnamefont {P.}~\bibnamefont
  {Metaxas}}, \bibinfo {author} {\bibfnamefont {J.}~\bibnamefont {Jamet}},
  \bibinfo {author} {\bibfnamefont {A.}~\bibnamefont {Mougin}}, \bibinfo
  {author} {\bibfnamefont {M.}~\bibnamefont {Cormier}}, \bibinfo {author}
  {\bibfnamefont {J.}~\bibnamefont {Ferr{\'e}}}, \bibinfo {author}
  {\bibfnamefont {V.}~\bibnamefont {Baltz}}, \bibinfo {author} {\bibfnamefont
  {B.}~\bibnamefont {Rodmacq}}, \bibinfo {author} {\bibfnamefont
  {B.}~\bibnamefont {Dieny}},\ and\ \bibinfo {author} {\bibfnamefont
  {R.}~\bibnamefont {Stamps}},\ }\href@noop {} {\bibfield  {journal} {\bibinfo
  {journal} {Physical review letters}\ }\textbf {\bibinfo {volume} {99}},\
  \bibinfo {pages} {217208} (\bibinfo {year} {2007})}\BibitemShut {NoStop}%
\bibitem [{\citenamefont {Rohart}\ and\ \citenamefont
  {Thiaville}(2013)}]{rohart2013skyrmion}%
  \BibitemOpen
  \bibfield  {author} {\bibinfo {author} {\bibfnamefont {S.}~\bibnamefont
  {Rohart}}\ and\ \bibinfo {author} {\bibfnamefont {A.}~\bibnamefont
  {Thiaville}},\ }\href@noop {} {\bibfield  {journal} {\bibinfo  {journal}
  {Physical Review B}\ }\textbf {\bibinfo {volume} {88}},\ \bibinfo {pages}
  {184422} (\bibinfo {year} {2013})}\BibitemShut {NoStop}%
\bibitem [{\citenamefont {Tejo}\ \emph {et~al.}(2018)\citenamefont {Tejo},
  \citenamefont {Riveros}, \citenamefont {Escrig}, \citenamefont {Guslienko},\
  and\ \citenamefont {Chubykalo-Fesenko}}]{tejo2018distinct}%
  \BibitemOpen
  \bibfield  {author} {\bibinfo {author} {\bibfnamefont {F.}~\bibnamefont
  {Tejo}}, \bibinfo {author} {\bibfnamefont {A.}~\bibnamefont {Riveros}},
  \bibinfo {author} {\bibfnamefont {J.}~\bibnamefont {Escrig}}, \bibinfo
  {author} {\bibfnamefont {K.}~\bibnamefont {Guslienko}},\ and\ \bibinfo
  {author} {\bibfnamefont {O.}~\bibnamefont {Chubykalo-Fesenko}},\ }\href@noop
  {} {\bibfield  {journal} {\bibinfo  {journal} {Scientific reports}\ }\textbf
  {\bibinfo {volume} {8}},\ \bibinfo {pages} {1} (\bibinfo {year}
  {2018})}\BibitemShut {NoStop}%
\bibitem [{\citenamefont {Tomasello}\ \emph {et~al.}(2018)\citenamefont
  {Tomasello}, \citenamefont {Guslienko}, \citenamefont {Ricci}, \citenamefont
  {Giordano}, \citenamefont {Barker}, \citenamefont {Carpentieri},
  \citenamefont {Chubykalo-Fesenko},\ and\ \citenamefont
  {Finocchio}}]{tomasello2018origin}%
  \BibitemOpen
  \bibfield  {author} {\bibinfo {author} {\bibfnamefont {R.}~\bibnamefont
  {Tomasello}}, \bibinfo {author} {\bibfnamefont {K.}~\bibnamefont
  {Guslienko}}, \bibinfo {author} {\bibfnamefont {M.}~\bibnamefont {Ricci}},
  \bibinfo {author} {\bibfnamefont {A.}~\bibnamefont {Giordano}}, \bibinfo
  {author} {\bibfnamefont {J.}~\bibnamefont {Barker}}, \bibinfo {author}
  {\bibfnamefont {M.}~\bibnamefont {Carpentieri}}, \bibinfo {author}
  {\bibfnamefont {O.}~\bibnamefont {Chubykalo-Fesenko}},\ and\ \bibinfo
  {author} {\bibfnamefont {G.}~\bibnamefont {Finocchio}},\ }\href@noop {}
  {\bibfield  {journal} {\bibinfo  {journal} {Physical Review B}\ }\textbf
  {\bibinfo {volume} {97}},\ \bibinfo {pages} {060402} (\bibinfo {year}
  {2018})}\BibitemShut {NoStop}%
\bibitem [{\citenamefont {Thiele}(1973)}]{thiele1973steady}%
  \BibitemOpen
  \bibfield  {author} {\bibinfo {author} {\bibfnamefont {A.}~\bibnamefont
  {Thiele}},\ }\href@noop {} {\bibfield  {journal} {\bibinfo  {journal}
  {Physical Review Letters}\ }\textbf {\bibinfo {volume} {30}},\ \bibinfo
  {pages} {230} (\bibinfo {year} {1973})}\BibitemShut {NoStop}%
\bibitem [{\citenamefont {Xing}\ \emph {et~al.}(2020)\citenamefont {Xing},
  \citenamefont {{\AA}kerman},\ and\ \citenamefont {Zhou}}]{xing2020enhanced}%
  \BibitemOpen
  \bibfield  {author} {\bibinfo {author} {\bibfnamefont {X.}~\bibnamefont
  {Xing}}, \bibinfo {author} {\bibfnamefont {J.}~\bibnamefont {{\AA}kerman}},\
  and\ \bibinfo {author} {\bibfnamefont {Y.}~\bibnamefont {Zhou}},\ }\href@noop
  {} {\bibfield  {journal} {\bibinfo  {journal} {Physical Review B}\ }\textbf
  {\bibinfo {volume} {101}},\ \bibinfo {pages} {214432} (\bibinfo {year}
  {2020})}\BibitemShut {NoStop}%
\bibitem [{\citenamefont {Yang}\ \emph {et~al.}(2022)\citenamefont {Yang},
  \citenamefont {Wu}, \citenamefont {Zhao}, \citenamefont {Liang},
  \citenamefont {Xia}, \citenamefont {Zhou}, \citenamefont {Xing},\ and\
  \citenamefont {Zhou}}]{yang2022inhibition}%
  \BibitemOpen
  \bibfield  {author} {\bibinfo {author} {\bibfnamefont {S.}~\bibnamefont
  {Yang}}, \bibinfo {author} {\bibfnamefont {K.}~\bibnamefont {Wu}}, \bibinfo
  {author} {\bibfnamefont {Y.}~\bibnamefont {Zhao}}, \bibinfo {author}
  {\bibfnamefont {X.}~\bibnamefont {Liang}}, \bibinfo {author} {\bibfnamefont
  {J.}~\bibnamefont {Xia}}, \bibinfo {author} {\bibfnamefont {Y.}~\bibnamefont
  {Zhou}}, \bibinfo {author} {\bibfnamefont {X.}~\bibnamefont {Xing}},\ and\
  \bibinfo {author} {\bibfnamefont {Y.}~\bibnamefont {Zhou}},\ }\href@noop {}
  {\bibfield  {journal} {\bibinfo  {journal} {Physical Review Applied}\
  }\textbf {\bibinfo {volume} {18}},\ \bibinfo {pages} {024030} (\bibinfo
  {year} {2022})}\BibitemShut {NoStop}%
\bibitem [{\citenamefont {Zhang}\ \emph {et~al.}(2015)\citenamefont {Zhang},
  \citenamefont {Zhao}, \citenamefont {Fangohr}, \citenamefont {Liu},
  \citenamefont {Xia}, \citenamefont {Xia},\ and\ \citenamefont
  {Morvan}}]{zhang2015skyrmion}%
  \BibitemOpen
  \bibfield  {author} {\bibinfo {author} {\bibfnamefont {X.}~\bibnamefont
  {Zhang}}, \bibinfo {author} {\bibfnamefont {G.}~\bibnamefont {Zhao}},
  \bibinfo {author} {\bibfnamefont {H.}~\bibnamefont {Fangohr}}, \bibinfo
  {author} {\bibfnamefont {J.~P.}\ \bibnamefont {Liu}}, \bibinfo {author}
  {\bibfnamefont {W.}~\bibnamefont {Xia}}, \bibinfo {author} {\bibfnamefont
  {J.}~\bibnamefont {Xia}},\ and\ \bibinfo {author} {\bibfnamefont
  {F.}~\bibnamefont {Morvan}},\ }\href@noop {} {\bibfield  {journal} {\bibinfo
  {journal} {Scientific reports}\ }\textbf {\bibinfo {volume} {5}},\ \bibinfo
  {pages} {1} (\bibinfo {year} {2015})}\BibitemShut {NoStop}%
\bibitem [{\citenamefont {Zhang}\ \emph
  {et~al.}(2017{\natexlab{b}})\citenamefont {Zhang}, \citenamefont
  {M{\"u}ller}, \citenamefont {Xia}, \citenamefont {Garst}, \citenamefont
  {Liu},\ and\ \citenamefont {Zhou}}]{zhang2017motion}%
  \BibitemOpen
  \bibfield  {author} {\bibinfo {author} {\bibfnamefont {X.}~\bibnamefont
  {Zhang}}, \bibinfo {author} {\bibfnamefont {J.}~\bibnamefont {M{\"u}ller}},
  \bibinfo {author} {\bibfnamefont {J.}~\bibnamefont {Xia}}, \bibinfo {author}
  {\bibfnamefont {M.}~\bibnamefont {Garst}}, \bibinfo {author} {\bibfnamefont
  {X.}~\bibnamefont {Liu}},\ and\ \bibinfo {author} {\bibfnamefont
  {Y.}~\bibnamefont {Zhou}},\ }\href@noop {} {\bibfield  {journal} {\bibinfo
  {journal} {New Journal of Physics}\ }\textbf {\bibinfo {volume} {19}},\
  \bibinfo {pages} {065001} (\bibinfo {year} {2017}{\natexlab{b}})}\BibitemShut
  {NoStop}%
\bibitem [{\citenamefont {Bo}\ \emph {et~al.}(2022{\natexlab{b}})\citenamefont
  {Bo}, \citenamefont {Hu}, \citenamefont {Kong}, \citenamefont {Ji},
  \citenamefont {Li}, \citenamefont {Zhang}, \citenamefont {Zhao},\ and\
  \citenamefont {Zhang}}]{bo2022velocity}%
  \BibitemOpen
  \bibfield  {author} {\bibinfo {author} {\bibfnamefont {L.}~\bibnamefont
  {Bo}}, \bibinfo {author} {\bibfnamefont {C.}~\bibnamefont {Hu}}, \bibinfo
  {author} {\bibfnamefont {L.}~\bibnamefont {Kong}}, \bibinfo {author}
  {\bibfnamefont {L.}~\bibnamefont {Ji}}, \bibinfo {author} {\bibfnamefont
  {Y.}~\bibnamefont {Li}}, \bibinfo {author} {\bibfnamefont {J.}~\bibnamefont
  {Zhang}}, \bibinfo {author} {\bibfnamefont {R.}~\bibnamefont {Zhao}},\ and\
  \bibinfo {author} {\bibfnamefont {X.}~\bibnamefont {Zhang}},\ }\href@noop {}
  {\bibfield  {journal} {\bibinfo  {journal} {Journal of Magnetism and Magnetic
  Materials}\ }\textbf {\bibinfo {volume} {546}},\ \bibinfo {pages} {168877}
  (\bibinfo {year} {2022}{\natexlab{b}})}\BibitemShut {NoStop}%
\bibitem [{\citenamefont {Walton}\ \emph {et~al.}(2013)\citenamefont {Walton},
  \citenamefont {Zeissler}, \citenamefont {Branford},\ and\ \citenamefont
  {Felton}}]{walton2013malts}%
  \BibitemOpen
  \bibfield  {author} {\bibinfo {author} {\bibfnamefont {S.~K.}\ \bibnamefont
  {Walton}}, \bibinfo {author} {\bibfnamefont {K.}~\bibnamefont {Zeissler}},
  \bibinfo {author} {\bibfnamefont {W.~R.}\ \bibnamefont {Branford}},\ and\
  \bibinfo {author} {\bibfnamefont {S.}~\bibnamefont {Felton}},\ }\href@noop {}
  {\bibfield  {journal} {\bibinfo  {journal} {IEEE Transactions on Magnetics}\
  }\textbf {\bibinfo {volume} {49}},\ \bibinfo {pages} {4795} (\bibinfo {year}
  {2013})}\BibitemShut {NoStop}%
\bibitem [{\citenamefont {McCray}\ \emph {et~al.}(2021)\citenamefont {McCray},
  \citenamefont {Cote}, \citenamefont {Li}, \citenamefont {Petford-Long},\ and\
  \citenamefont {Phatak}}]{mccray2021understanding}%
  \BibitemOpen
  \bibfield  {author} {\bibinfo {author} {\bibfnamefont {A.~R.}\ \bibnamefont
  {McCray}}, \bibinfo {author} {\bibfnamefont {T.}~\bibnamefont {Cote}},
  \bibinfo {author} {\bibfnamefont {Y.}~\bibnamefont {Li}}, \bibinfo {author}
  {\bibfnamefont {A.~K.}\ \bibnamefont {Petford-Long}},\ and\ \bibinfo {author}
  {\bibfnamefont {C.}~\bibnamefont {Phatak}},\ }\href@noop {} {\bibfield
  {journal} {\bibinfo  {journal} {Physical Review Applied}\ }\textbf {\bibinfo
  {volume} {15}},\ \bibinfo {pages} {044025} (\bibinfo {year}
  {2021})}\BibitemShut {NoStop}%
\bibitem [{\citenamefont {Tai}\ \emph {et~al.}(2018)\citenamefont {Tai},
  \citenamefont {Smalyukh} \emph {et~al.}}]{tai2018static}%
  \BibitemOpen
  \bibfield  {author} {\bibinfo {author} {\bibfnamefont {J.-S.~B.}\
  \bibnamefont {Tai}}, \bibinfo {author} {\bibfnamefont {I.~I.}\ \bibnamefont
  {Smalyukh}}, \emph {et~al.},\ }\href@noop {} {\bibfield  {journal} {\bibinfo
  {journal} {Physical Review Letters}\ }\textbf {\bibinfo {volume} {121}},\
  \bibinfo {pages} {187201} (\bibinfo {year} {2018})}\BibitemShut {NoStop}%
\bibitem [{\citenamefont {Voinescu}\ \emph {et~al.}(2020)\citenamefont
  {Voinescu}, \citenamefont {Tai}, \citenamefont {Smalyukh} \emph
  {et~al.}}]{voinescu2020hopf}%
  \BibitemOpen
  \bibfield  {author} {\bibinfo {author} {\bibfnamefont {R.}~\bibnamefont
  {Voinescu}}, \bibinfo {author} {\bibfnamefont {J.-S.~B.}\ \bibnamefont
  {Tai}}, \bibinfo {author} {\bibfnamefont {I.~I.}\ \bibnamefont {Smalyukh}},
  \emph {et~al.},\ }\href@noop {} {\bibfield  {journal} {\bibinfo  {journal}
  {Physical Review Letters}\ }\textbf {\bibinfo {volume} {125}},\ \bibinfo
  {pages} {057201} (\bibinfo {year} {2020})}\BibitemShut {NoStop}%
\bibitem [{\citenamefont {Bo}\ \emph {et~al.}(2021)\citenamefont {Bo},
  \citenamefont {Ji}, \citenamefont {Hu}, \citenamefont {Zhao}, \citenamefont
  {Li}, \citenamefont {Zhang},\ and\ \citenamefont {Zhang}}]{bo2021spin}%
  \BibitemOpen
  \bibfield  {author} {\bibinfo {author} {\bibfnamefont {L.}~\bibnamefont
  {Bo}}, \bibinfo {author} {\bibfnamefont {L.}~\bibnamefont {Ji}}, \bibinfo
  {author} {\bibfnamefont {C.}~\bibnamefont {Hu}}, \bibinfo {author}
  {\bibfnamefont {R.}~\bibnamefont {Zhao}}, \bibinfo {author} {\bibfnamefont
  {Y.}~\bibnamefont {Li}}, \bibinfo {author} {\bibfnamefont {J.}~\bibnamefont
  {Zhang}},\ and\ \bibinfo {author} {\bibfnamefont {X.}~\bibnamefont {Zhang}},\
  }\href@noop {} {\bibfield  {journal} {\bibinfo  {journal} {Applied Physics
  Letters}\ }\textbf {\bibinfo {volume} {119}},\ \bibinfo {pages} {212408}
  (\bibinfo {year} {2021})}\BibitemShut {NoStop}%
\bibitem [{\citenamefont {Pollard}\ \emph {et~al.}(2017)\citenamefont
  {Pollard}, \citenamefont {Garlow}, \citenamefont {Yu}, \citenamefont {Wang},
  \citenamefont {Zhu},\ and\ \citenamefont {Yang}}]{pollard2017observation}%
  \BibitemOpen
  \bibfield  {author} {\bibinfo {author} {\bibfnamefont {S.~D.}\ \bibnamefont
  {Pollard}}, \bibinfo {author} {\bibfnamefont {J.~A.}\ \bibnamefont {Garlow}},
  \bibinfo {author} {\bibfnamefont {J.}~\bibnamefont {Yu}}, \bibinfo {author}
  {\bibfnamefont {Z.}~\bibnamefont {Wang}}, \bibinfo {author} {\bibfnamefont
  {Y.}~\bibnamefont {Zhu}},\ and\ \bibinfo {author} {\bibfnamefont
  {H.}~\bibnamefont {Yang}},\ }\href@noop {} {\bibfield  {journal} {\bibinfo
  {journal} {Nature communications}\ }\textbf {\bibinfo {volume} {8}},\
  \bibinfo {pages} {1} (\bibinfo {year} {2017})}\BibitemShut {NoStop}%
\bibitem [{\citenamefont {He}\ \emph {et~al.}(2017)\citenamefont {He},
  \citenamefont {Peng}, \citenamefont {Zhu}, \citenamefont {Li}, \citenamefont
  {Cai}, \citenamefont {Li}, \citenamefont {Wei}, \citenamefont {Gu},
  \citenamefont {Wang}, \citenamefont {Zhao} \emph
  {et~al.}}]{he2017realization}%
  \BibitemOpen
  \bibfield  {author} {\bibinfo {author} {\bibfnamefont {M.}~\bibnamefont
  {He}}, \bibinfo {author} {\bibfnamefont {L.}~\bibnamefont {Peng}}, \bibinfo
  {author} {\bibfnamefont {Z.}~\bibnamefont {Zhu}}, \bibinfo {author}
  {\bibfnamefont {G.}~\bibnamefont {Li}}, \bibinfo {author} {\bibfnamefont
  {J.}~\bibnamefont {Cai}}, \bibinfo {author} {\bibfnamefont {J.}~\bibnamefont
  {Li}}, \bibinfo {author} {\bibfnamefont {H.}~\bibnamefont {Wei}}, \bibinfo
  {author} {\bibfnamefont {L.}~\bibnamefont {Gu}}, \bibinfo {author}
  {\bibfnamefont {S.}~\bibnamefont {Wang}}, \bibinfo {author} {\bibfnamefont
  {T.}~\bibnamefont {Zhao}}, \emph {et~al.},\ }\href@noop {} {\bibfield
  {journal} {\bibinfo  {journal} {Applied Physics Letters}\ }\textbf {\bibinfo
  {volume} {111}},\ \bibinfo {pages} {202403} (\bibinfo {year}
  {2017})}\BibitemShut {NoStop}%
\bibitem [{\citenamefont {Lin}\ \emph {et~al.}(2018)\citenamefont {Lin},
  \citenamefont {Liu}, \citenamefont {Poellath}, \citenamefont {Zhang},
  \citenamefont {Ji}, \citenamefont {Lei}, \citenamefont {Yun}, \citenamefont
  {Xi}, \citenamefont {Yang}, \citenamefont {Xing} \emph
  {et~al.}}]{lin2018observation}%
  \BibitemOpen
  \bibfield  {author} {\bibinfo {author} {\bibfnamefont {T.}~\bibnamefont
  {Lin}}, \bibinfo {author} {\bibfnamefont {H.}~\bibnamefont {Liu}}, \bibinfo
  {author} {\bibfnamefont {S.}~\bibnamefont {Poellath}}, \bibinfo {author}
  {\bibfnamefont {Y.}~\bibnamefont {Zhang}}, \bibinfo {author} {\bibfnamefont
  {B.}~\bibnamefont {Ji}}, \bibinfo {author} {\bibfnamefont {N.}~\bibnamefont
  {Lei}}, \bibinfo {author} {\bibfnamefont {J.}~\bibnamefont {Yun}}, \bibinfo
  {author} {\bibfnamefont {L.}~\bibnamefont {Xi}}, \bibinfo {author}
  {\bibfnamefont {D.}~\bibnamefont {Yang}}, \bibinfo {author} {\bibfnamefont
  {T.}~\bibnamefont {Xing}}, \emph {et~al.},\ }\href@noop {} {\bibfield
  {journal} {\bibinfo  {journal} {Physical Review B}\ }\textbf {\bibinfo
  {volume} {98}},\ \bibinfo {pages} {174425} (\bibinfo {year}
  {2018})}\BibitemShut {NoStop}%
\bibitem [{\citenamefont {Bateman}\ and\ \citenamefont
  {Erd{\"o}s}(1951)}]{bateman1951geometrical}%
  \BibitemOpen
  \bibfield  {author} {\bibinfo {author} {\bibfnamefont {P.}~\bibnamefont
  {Bateman}}\ and\ \bibinfo {author} {\bibfnamefont {P.}~\bibnamefont
  {Erd{\"o}s}},\ }\href@noop {} {\bibfield  {journal} {\bibinfo  {journal} {The
  American Mathematical Monthly}\ }\textbf {\bibinfo {volume} {58}},\ \bibinfo
  {pages} {306} (\bibinfo {year} {1951})}\BibitemShut {NoStop}%
\bibitem [{\citenamefont {Anjan}\ \emph {et~al.}(2017)\citenamefont {Anjan},
  \citenamefont {Soumyanarayanan}, \citenamefont {Raju}, \citenamefont {L},
  \citenamefont {Oyarce}, \citenamefont {Anthony}, \citenamefont {Tan},
  \citenamefont {Mi-Young}, \citenamefont {Im},\ and\ \citenamefont
  {P}}]{Anjan2017Tunable}%
  \BibitemOpen
  \bibfield  {author} {\bibinfo {author} {\bibnamefont {Anjan}}, \bibinfo
  {author} {\bibnamefont {Soumyanarayanan}}, \bibinfo {author} {\bibnamefont
  {Raju}}, \bibinfo {author} {\bibfnamefont {A.}~\bibnamefont {L}}, \bibinfo
  {author} {\bibfnamefont {G.}~\bibnamefont {Oyarce}}, \bibinfo {author}
  {\bibfnamefont {K.~C.}\ \bibnamefont {Anthony}}, \bibinfo {author}
  {\bibnamefont {Tan}}, \bibinfo {author} {\bibnamefont {Mi-Young}}, \bibinfo
  {author} {\bibnamefont {Im}},\ and\ \bibinfo {author} {\bibfnamefont
  {A.}~\bibnamefont {P}},\ }\href@noop {} {\bibfield  {journal} {\bibinfo
  {journal} {Nature materials}\ } (\bibinfo {year} {2017})}\BibitemShut
  {NoStop}%
\bibitem [{\citenamefont {Suess}\ \emph {et~al.}(2018)\citenamefont {Suess},
  \citenamefont {Vogler}, \citenamefont {Bruckner}, \citenamefont
  {Heistracher},\ and\ \citenamefont {Abert}}]{2018A}%
  \BibitemOpen
  \bibfield  {author} {\bibinfo {author} {\bibfnamefont {D.}~\bibnamefont
  {Suess}}, \bibinfo {author} {\bibfnamefont {C.}~\bibnamefont {Vogler}},
  \bibinfo {author} {\bibfnamefont {F.}~\bibnamefont {Bruckner}}, \bibinfo
  {author} {\bibfnamefont {P.}~\bibnamefont {Heistracher}},\ and\ \bibinfo
  {author} {\bibfnamefont {C.}~\bibnamefont {Abert}},\ }\href@noop {}
  {\bibfield  {journal} {\bibinfo  {journal} {Aip Advances}\ }\textbf {\bibinfo
  {volume} {8}} (\bibinfo {year} {2018})}\BibitemShut {NoStop}%
\bibitem [{\citenamefont {Yin}\ \emph {et~al.}(2016)\citenamefont {Yin},
  \citenamefont {Li}, \citenamefont {Kong}, \citenamefont {Lake}, \citenamefont
  {Chien},\ and\ \citenamefont {Zang}}]{2016Topological}%
  \BibitemOpen
  \bibfield  {author} {\bibinfo {author} {\bibfnamefont {G.}~\bibnamefont
  {Yin}}, \bibinfo {author} {\bibfnamefont {Y.}~\bibnamefont {Li}}, \bibinfo
  {author} {\bibfnamefont {L.}~\bibnamefont {Kong}}, \bibinfo {author}
  {\bibfnamefont {R.~K.}\ \bibnamefont {Lake}}, \bibinfo {author}
  {\bibfnamefont {C.~L.}\ \bibnamefont {Chien}},\ and\ \bibinfo {author}
  {\bibfnamefont {J.}~\bibnamefont {Zang}},\ }\href@noop {} {\bibfield
  {journal} {\bibinfo  {journal} {Physical Review B}\ }\textbf {\bibinfo
  {volume} {93}} (\bibinfo {year} {2016})}\BibitemShut {NoStop}%
\end{thebibliography}%
	
\end{document}